\documentclass[11pt,a4paper]{article}
\usepackage{longtable}
\usepackage{amsmath}
\usepackage{amssymb}
\usepackage{graphicx}
\usepackage{bm}
\usepackage{footmisc}

\newcommand*{\La}{\cal{L}}
\newcommand*{\no}{\noindent}
\newcommand*{\bea}{\begin{eqnarray}}
\newcommand*{\eea}{\end{eqnarray}}
\newcommand*{\be}{\begin{equation}}
\newcommand*{\ee}{\end{equation}}
\newcommand*{\pd}{\partial}
\newcommand*{\pdm}{\pd_{\mu}}
\newcommand*{\pdn}{\pd_{\nu}}

\newcommand*{\pref}[1]{(\ref{#1})}

\newcommand*{\mn}{{\mu\nu}}

\newcommand*{\prefr}[2]{(\ref{#1}-\ref{#2})} 
\newcommand*{\nn}{\nonumber}

\newcommand*{\tr}{\mathrm{tr}}

\newcommand*{\tl}{\mathrm{tl}}

\newcommand{\bma}{\begin{pmatrix}}
\newcommand{\ema}{\end{pmatrix}}

\usepackage[square,comma,sort&compress,numbers]{natbib}

\title{Two- and three-point functions in Landau gauge Yang-Mills-Higgs theory}

\author{Axel Maas, Tajdar Mufti\\
Institute for Theoretical Physics, Friedrich-Schiller-University Jena,\\
Max-Wien-Platz 1, D-07743 Jena, Germany}

\begin{document}

\maketitle

\begin{abstract}
Yang-Mills-Higgs theory offers a rich set of physics. In particular, in some region of its parameter space it has QCD-like behavior, while in some other range it is Higgs-like. Furthermore, for the choice of the gauge group SU(2) and an SU(2) Higgs flavor symmetry it is the Higgs sector of the standard model. Therefore, it is possible to study a plethora of phenomena within a single theory. Here the standard-model version is studied using lattice gauge theory. Choosing non-aligned minimal Landau gauge, its propagators and three-point vertices will be determined in both the QCD-like and Higgs-like domains. This permits to test various proposals for how confinement works, as well as how confinement and the Higgs effect differ. The correlations functions are found to exhibit a different behavior, depending on whether the lowest mass scalar flavor singlet is lighter than the vector triplet, heavier and stable, or unstable against decay into two vector triplets.
\end{abstract}

\section{Introduction}

Combining a non-Abelian gauge theory with a fundamental scalar, called here for convenience the Higgs, yields a theory which offers a plethora of interesting phenomena. In particular, with two Higgs flavors (a complex doublet) in combination with an SU(2) gauge group, it forms the Higgs sector of the standard model. However, without QED or other custodial symmetry breaking effects, all gauge bosons are degenerate, and will therefore be referred to as $W$. On the other hand, if the Higgs effect should not be operative, the theory should exhibit a QCD-like behavior, and especially confinement. This theory is therefore an excellent laboratory to understand both types of physics, and especially how they differ.

The first lesson about their relation has been learned already long ago \cite{Fradkin:1978dv}: When regulated with a lattice cutoff, there is no physical distinction between both phases, and any point in the quantum phase diagram is connected with any other analytically. This has been confirmed in a multitude of lattice simulations, see especially \cite{Bonati:2009pf,Caudy:2007sf,Jersak:1985nf,Langguth:1985eu}. However, it is not yet clear whether such a theory may be trivial \cite{Callaway:1988ya}, and therefore whether this statement is regulator-dependent. In the context of the standard model, this problem may either be alleviated by dynamical effects in the interplay of all sectors \cite{Gies:2013pma,Callaway:1988ya} or by new physics. Here, the precise resolution of this problem is of no interest, and we use the lattice cutoff as a convenient way to encode any of these effects, under the assumption that this will not severely affect the low-energy physics, i.\ e.\ below 1 TeV, in which we are interested here. The question of how we then characterize both regimes will be detailed in section \ref{sphase} below, and will be more pragmatic than fundamental.

However, this coincidence is only necessary for observables. Gauge-de\-pen\-dent quantities, and especially propagators and vertices, can exhibit in suitable gauges a qualitative difference \cite{Caudy:2007sf,Greensite:2008ss,Greensite:2004ke}. But also the confinement mechanism and the Higgs mechanism with its gauge-dependent vacuum expectation value\footnote{Which is also the reason why the Higgs phase is perturbatively only accessible in some gauges \cite{Lee:1974zg}.} \cite{Caudy:2007sf,Frohlich:1981yi,Frohlich:1980gj,Maas:2012ct} are very likely gauge-dependent \cite{Kugo:1979gm}. Thus the study of correlation functions can serve as a valuable tool in understanding these mechanisms, as has been done for Yang-Mills theory \cite{Maas:2011se}. Especially, several predictions and functional results are available for the present case \cite{Gies:2013pma,Fister:2010yw,Macher:2011ad,Fischer:2009tn,Hopfer:2013via,Mitter:2013me,Capri:2012ah}, and therefore it is worthwhile to check them explicitly. Furthermore, these correlation functions provide a valuable input and cross-check for other methods, e.\ g.\ functional methods \cite{Maas:2011se}. They therefore represent quantities of interest in themselves.

Of course, since especially confinement is non-per\-tur\-ba\-tive, non-per\-tur\-ba\-tive methods are necessary to determine the correlation functions. For this purpose, lattice gauge theory will be employed here. The technical details are given in section \ref{stech}. In addition, the general setup of the theory is briefly given in continuum terms in section \ref{ssetup}. Results for the propagators, both in position and momentum space, are presented in section \ref{sprop}, and for the three-point vertices in section \ref{svertex}. Since already the three-point vertices challenged our computational resources to the utmost, the ever more demanding four-point vertices were beyond our reach at the current time, see section \ref{s4p}. A brief summary and some conclusions are given in section \ref{sconclusions}. Some technical details and general comments are relegated to the appendices.

This work extends the previous results \cite{Maas:2012tj,Maas:2012zf,Mufti:2013uxa}. There will also appear a companion paper soon, which addresses certain gauge-invariant aspects of the physics of this theory \cite{Maas:unpublished3}, which results will only be stated here. The results here can also be compared to the quenched case for the scalar sector \cite{Maas:2011yx} or the gauge sector \cite{Maas:2011se}.

\section{Technical details}\label{stech}

\subsection{Setup}\label{ssetup}

The theory to be investigated here is two flavors of scalar particles $\phi$ coupled to a non-Abelian gauge field $W$, with the (Euclidean) action
\bea
{\La}&=&-\frac{1}{4}W_\mn^aW^\mn_a+(D_\mu\phi)^\dagger D^\mu\phi-\gamma(\phi\phi^\dagger)^2-\frac{m_0^2}{2}\phi\phi^\dagger\label{action}\\
W_\mn^a&=&\pdm W_\nu^a-\pdn W_\mu^a-g f^{abc} W_\mu^b W_\nu^c\nn\\
D_\mu^{ij}&=&\pd_\mu\delta^{ij}-igW_\mu^a\tau^{ij}_a\nn,
\eea
\no where $g$ is the gauge-coupling, $\gamma$ and $m_0$ the parameters of the Higgs potential, and $f^{abc}$ and $\tau^a$ are the structure constants and generators of the gauge group, respectively. The gauge group is chosen to be the weak isospin gauge group SU(2). Thus, the complex doublet $\phi$ contains four real scalar degrees of freedom, exhibiting an SU(2) custodial symmetry, which is in fact just the (Higgs-)flavor symmetry. The Lagrangian is invariant under the latter symmetry, as an explicit flavor-symmetry-breaking term is absent. This symmetry is also found to be not broken spontaneously for any of the parameters to be simulated here. It will therefore be repeatedly convenient to employ the notation \cite{Shifman:2012zz}
\be
X=\bma \phi_1 & -\phi_2^* \cr \phi_2 & \phi_1^* \ema=\phi^\dagger_i\phi_i\varphi\label{x}
\ee
\no which makes this fact explicit: Gauge transformations act on this matrix as a left multiplication, while flavor transformations act as a right multiplication. As given by the second equality, this can be written as the length of the Higgs field multiplied by an SU(2)-valued matrix $\varphi$.

It is important to make a remark here concerning the naming conventions. In this work, we will adhere strictly to the above prescribed naming scheme of calling the (gauge-dependent) elementary fields Higgs and $W$, in accordance with the PDG \cite{pdg}, and the phenomenological language. In contrast, based on the works \cite{Banks:1979fi,'tHooft:1979bj,Frohlich:1980gj,Frohlich:1981yi}, certain gauge-invariant composite operators have in the lattice literature been denoted as Higgs and $W$ boson, for reasons discussed in \cite{Maas:2012tj} and in section \ref{sphase}. Thus, one should be wary when comparing these different resources.

The aim here are the gauge-dependent two-point and three-point functions. Hence, it is necessary to fix a gauge. For this the Landau gauge $\pdm W_\mu^a=0$ will be chosen, which requires in the continuum to add a ghost field $c$ and an anti-ghost field $\bar c$ with the Lagrangian
\bea
{\La}_g&=&\bar{c}_a\pdm D_\mu^{ab} c_b\nn\\
D_\mu^{ab}&=&\pd_\mu\delta^{ab}+g f^{ab}_c W_\mu^c\nn.
\eea
\no However, this does not yet specify the gauge completely. First of all, due to the Gribov-Singer ambiguity \cite{Maas:2011se,Gribov:1977wm,Singer:1978dk}, this is only a perturbative definition, which requires a non-perturbative extension to make it well-defined. For this purpose, the minimal Landau gauge prescription will be used, i.\ e.\ an average over all gauge copies satisfying the Landau gauge condition, for which also the Faddeev-Popov operator $-\pdm D_\mu^{ab}$ is positive semi-definite, will be performed \cite{Maas:2011se}. Exploratory investigations \cite{Maas:2010nc} indicate that, as in Yang-Mills theory \cite{Maas:2011se}, alternative choices do have some influence on the propagators, and likely the vertices. Thus, it is important to only compare minimal Landau gauge results with each other.

Second, this does not specify how to deal with the global part of the gauge symmetry. However, this is required in presence of a Higgs effect. A convenient choice is a non-aligned gauge, i.\ e.\ one in which the global gauge degree of freedom is integrated over \cite{Maas:2012ct}. This implies that the space-time average of the Higgs field, and any other space-time-independent quantity with a gauge-index, is vanishing for every configuration identically. This especially implies that the Higgs expectation value is zero. This gauge has a number of advantages. Foremost, it is also a well-defined gauge choice even when the Higgs phase is not operative, and can therefore be defined throughout the whole phase diagram. Secondly, it is also technically advantageous \cite{Maas:2012ct}: On the one hand it reduces the number of non-vanishing color tensors to the minimal one. Secondly, it reduces in lattice calculations the statistical noise, since many disconnected contributions vanish.

Of course, such a gauge choice implies that a perturbative treatment is not trivially possible, as to all orders in perturbation theory the gauge bosons will remain massless. However, for certain quantities it is still possible to compare to perturbative results in other gauges. This is detailed in appendix \ref{a:pt}.

Other gauge choices are of course also possible. However, this choice yields the lowest number of independent tensor structures with the simplest renormalization structure, and is applicable throughout the phase diagram. It is also the one used in the functional calculations outside the Higgs regime \cite{Fister:2010yw,Macher:2011ad,Fischer:2009tn,Hopfer:2013via,Mitter:2013me}. Hence, it will be used here. Some aspects of alternative choices are discussed in appendix \ref{a:remarks}.

\subsection{Creation of configurations}\label{s:configs}

The lattice calculations presented use the techniques described in \cite{Cucchieri:2006tf,Maas:2010nc,Maas:2012tj}. For the sake of completeness, the details will be repeated here.

The starting point is the unimproved lattice version of the action \pref{action}, given by \cite{Montvay:1994cy},
\bea
S&=&\beta\sum_x\Big(1-\frac{1}{2}\sum_{\mu<\nu}\Re\tr U_\mn(x)+\phi^\dagger(x)\phi(x)+\lambda\left(\phi(x)^\dagger\phi(x)-1\right)^2\nonumber\\
&&-\kappa\sum_\mu\Big(\phi(x)^\dagger U_\mu(x)\phi(x+e_\mu)+\phi(x+e_\mu)^\dagger U_\mu(x)^\dagger\phi(x)\Big)\Big)\label{laction}\\
U_\mn(x)&=&U_\mu(x)U_\nu(x+e_\mu)U_\mu(x+e_\nu)^{+}U_\nu(x)^{+}\label{plaq}\\
W_\mu&=&\frac{1}{2agi}(U_\mu(x)-U_\mu(x)^\dagger)+{\cal O}(a^2)\label{wdef}\\
\beta&=&\frac{4}{g^2}\nn\\
a^2m_0^2&=&\frac{(1-2\lambda)}{\kappa}-8\nn\\
\lambda&=&\kappa^2\gamma\nn.
\eea
\no In this expression $a$ is the lattice spacing, $U_\mu$ the link variable $\exp(iga W_\mu)$, $\phi$ again the Higgs field, the bare lattice couplings depend on the bare continuum couplings in the described way, and $e_\mu$ is the unit vector in $\mu$ direction. They are therefore the couplings at the lattice cut-off, which is essentially given by $1/a$, with the largest energy accessible being $4/a$, corresponding to a momentum across the body-diagonal of the cubic lattice of extension $N$ in each direction.

Choosing a physical scale is not an entirely trivial issue \cite{Maas:2013eh,Maas:unpublished3}, especially when a consistent scale setting between the Higgs and the confinement region should be achieved. To circumvent this in a constructive way, here the lighter of the masses of the ground states in the $0^{+}$ flavor singlet and $1^{-}$ flavor triplet channels, obtained with the methods described below in section \ref{s:bs}, will be set to 80.375 GeV. This gives for a light Higgs in the would-be Higgs phase the experimentally observed $W$ mass. This will be discussed further in section \ref{sphase}. The set of lattice parameters used for most of the calculations is given in table \ref{configs} below.

The generation of configurations follows \cite{Maas:2010nc}, using a combination of one heat-bath and five over-relaxation sweeps for the gauge fields according to \cite{Cucchieri:2006tf}, and in between each of these 6 sweeps of the gauge fields one Metropolis sweep for the Higgs field using a Gaussian proposal. The width of the proposal is adaptively tuned to achieve a 50\% acceptance probability. This should balance the movement through configuration space compared to the finding of relevant configurations. These updates have been performed lexicographical. These 12 sweeps together constitute a single update for the field configuration. The auto-correlation time of the plaquette is of the order of 1 or less such update. Thus, $N$ such updates separate a measurement of a gauge-invariant observable, to reduce the auto-correlation time. Because of the gauge-dependency and the issue of finding the same Gribov copy, the gauge-dependent quantities determined here have not been used to determine the auto-correlation time. For the thermalization, $2(10N+300)$ such updates have been performed.  Furthermore, all calculations involved many independent runs, to further reduce correlations.

All errors have been calculated using bootstrap with 1000 re-samplings and give a, possibly asymmetric, 67.5\% interval, i.\ e.\ approximately 1$\sigma$ interval.

The code, including the one to determine the bound states in section \ref{s:bs}, has been checked by comparing to the results in \cite{Langguth:1985dr,Jersak:1985nf}. The code for gauge-fixing and the pure gauge propagators and vertices has been extensively tested in the Yang-Mills case. The code for the correlation functions involving matter fields has been implemented independently twice.

\subsection{Gauge fixing}

To obtain the gauge-dependent correlation functions, a subset of the configurations are gauge fixed to the non-aligned minimal Landau (NML) gauge. Because the  gauge-fixing itself tends to show a longer auto-correlation time than the plaquette \cite{Maas:2008ri}, at least $2(N+30)$ updates have been performed between measurements of gauge-fixed quantities. Furthermore, since the relation \pref{wdef} only holds for a positive Polyakov loop \cite{Karsch:1994xh}, configurations with negative Polyakov loop in any direction have not been included for gauge-fixed measurements\footnote{Because the Higgs field explicitly breaks the center symmetry, this could not be solved by a center transformation. However, the value of the Polyakov loops are rather small, and thus the effect should be minor \cite{Maas:unpublished}.}.

The local part of the gauge-fixing has then been performed using a self-tuning stochastic over-relaxation algorithm with a quality parameter $e_6$ smaller than $10^{-12}$, see \cite{Cucchieri:2006tf} for details. This also automatically yields a gauge copy with positive semi-definite Faddeev-Popov operator. Since there appears to be no bias in the selection of which gauge copy is obtained \cite{Maas:2011se,Maas:2011ba,Maas:2013vd}, taking just this so created random one is equivalent to averaging over this set after ensemble averaging, thus implementing minimal Landau gauge \cite{Maas:2013vd}.

To implement the non-alignment a random global gauge transformation was performed after the fixing to minimal Landau gauge \cite{Maas:2012ct}. This only ensures the vanishing of the Higgs expectation value and similar quantities on the average, instead of for any configuration individually, but the additional noise is out-weighted by the gain in statistics of independent configurations. In fact, for typical lattice settings with physics similar to the standard model, the fluctuations around the average length of the Higgs field is very small, and only increases slightly for the most extreme cases investigated here. Thus, this a small effect.

\subsection{Propagators}\label{s:propagators}

The propagators have been obtained with the methods described in \cite{Maas:2010nc,Cucchieri:2006tf}. Using the definition \pref{wdef} for the $W$ field, the $W$ propagator is given by \cite{Cucchieri:2006tf}
\be
D_\mn^{ab}=\langle W_\mu^a W_\nu^b\rangle\nn.
\ee
\no In NML gauge it is transverse and color-diagonal with a single dressing function $Z$, and multiplicatively renormalized with the wave-function renormalization factor $Z_W$
\be
D_\mn^{ab}=\delta^{ab}\left(\delta_\mn-\frac{p_\mu p_\nu}{p^2}\right)\frac{Z_WZ(p^2)}{p^2}\nn.
\ee
\no The renormalization scheme is to demand $Z_WZ(\mu^2)=1$, which can be used irrespective of the phase diagram region, as long as $\mu\neq 0$.

The ghost propagator is considerably more complicated than the $W$, as on the lattice it is given as an inverse of the Faddeev-Popov operator \cite{Zwanziger:1993dh},
\be
D_G^{ab}(p)=\frac{1}{V}\left\langle(-\pdm D_\mu^{ab})^{-1}(p)\right\rangle\nn.
\ee
\no The expression for the Faddeev-Popov operator is lengthy, and can be found in \cite{Maas:2011se,Zwanziger:1993dh}. Remaining with a non-aligned gauge, it is required to invert this operator on the sub-space orthogonal to constant modes, as the latter correspond to global color rotations. Since the Faddeev-Popov operator is positive\footnote{On a finite lattice there are no additional zero modes.} and symmetric, this inversion is done on a point-source using a conjugate gradient algorithm, see \cite{Cucchieri:2006tf}. The resulting propagator is color-diagonal and thus has a single dressing function $G$, and is also renormalized multiplicatively,
\be
D_G^{ab}=-\delta^{ab}\frac{Z_GG(p^2)}{p^2}\nn,
\ee
\no where the same renormalization condition will be used as for the $W$ propagator.

It should be noted that the two renormalization constants for the $W$ and ghost propagators are not independent, and are linked by the condition $Z_{\bar{c}cW}=Z_W Z_G^2$ to the renormalization constant of the ghost-$W$ vertex $Z_{\bar{c}cW}$. Since the latter can be chosen to be one \cite{Taylor:1971ff,vonSmekal:2009ae}, the renormalization constants are then uniquely linked, up to lattice artifacts, in this so-called miniMOM scheme.

The Higgs propagator is the most straight-forward one, given by
\be
D_H^{ab}=\langle\phi(p)^{a\dagger}\phi(p)^{b}\rangle\nn.
\ee
\no However, the renormalization is more involved \cite{Bohm:2001yx}. The Higgs propagator requires besides the multiplicative wave-function renormalization also an additive mass renormalization. The renormalized propagator is given by
\be
D_H^{ab}(p^2)=\frac{\delta^{ab}}{Z_H(p^2+m^2)+\Pi_H(p^2)+\delta m^2}\nn,
\ee
\no where $\Pi_H$ is its self-energy, and $Z_H$ and $\delta m^2$ are the wave-function and mass renormalization constants, respectively. The two renormalization conditions implemented are \cite{Maas:2010nc}
\bea
D_H^{ab}(\mu^2)&=&\frac{\delta^{ab}}{\mu^2+m_H^2}\label{scheme1}\\
\left.\frac{\pd D_H^{ab}(p^2)}{\pd |p|}\right|_{|p|=\mu}&=&-\frac{2\mu\delta^{ab}}{(\mu^2+m_H^2)^2}\label{scheme2},
\eea
\no with $m_H=\mu$. Selecting $\mu$, $m_H$, and alongside the two conditions for the ghost and $W$ propagator, thus defines our (mass-dependent) renormalization scheme. 

For both the $W$ and the Higgs it will be interesting to also calculate the propagator in position space, the so-called Schwinger function. It is obtained from the renormalized momentum-space propagators $D$ as
\be
\Delta(t)=\frac{1}{a\pi}\frac{1}{N}\sum_{P_0=0}^{N_t-1}\cos\left(\frac{2\pi tP_0}{N_t}\right)D(P_0^2)\label{schwinger}.
\ee
\no Note that the additive renormalization for the Higgs makes it much easier to calculate the momentum-space propagator first and then afterwards this position-space function, while this is not relevant for the $W$ propagator.

It is furthermore important to note that all the propagators presented here are diagonal in color and flavor space. Thus, they are independent under global color and flavor rotations, see also appendix \ref{a:pt}. Especially this implies that their traces are identical to the ones in an aligned Landau gauge, e.\ g.\ the Landau-gauge limit of 't Hooft gauges, even though the individual color components no longer coincide as they do in the present non-aligned gauge.

The results for the propagators will be shown with momenta selected along several different directions, including edge, space, and space-time diagonals \cite{Cucchieri:2006tf}, to permit assessment of the impact of violation of rotational symmetry. The effects turn out to be small, and of little relevance for the findings here. Furthermore, for particular calculations, e.\ g.\ like the Schwinger function \pref{schwinger}, momenta are selected which are least sensitive to rotational symmetry violations at long distances, i.\ e.\ edge momenta.

\subsection{Vertices}

As noted in the introduction, only the three-point vertices were statistically feasible at the current time. Of these, there are three in the NML gauge. These are the ghost-$W$ vertex, the three-$W$ vertex, and the Higgs-$W$ vertex. While their determination is straight-forward \cite{Maas:2011se,Cucchieri:2006tf}, there are a number of subtleties concerning their tensor structures to be taken care of.

Three-point functions can have various tensor structures. Since only the non-amputated full correlation functions can be obtained in lattice calculations, it is necessary to isolate the various tensor structures. The choice of a non-aligned gauge makes for all three-point functions the connected and disconnected part coincide. Furthermore, to determine a normalized dressing function $A$ of a tensor structure from a connected three-point expectation value $G$, the simplest prescription is the projection \cite{Cucchieri:2006tf,Maas:2011se}
\be
A=\frac{\Gamma_{ijk} G_{ijk}}{\Gamma_{abc} D_{ad}D_{be}D_{cf} \Gamma_{def}}\nn,
\ee
\no where $\Gamma$ is some tensor structure, and the indices are generic multi-indices for internal and Lorentz degrees of freedom. A judicious choice are tensor structures which either coincide with the tree-level one, or are orthogonal to it. Then, this expression is one or zero, if the dressing function coincides with the tree-level one. The $D$ are symbolically the propagators of the three legs, and including them amputates the result. On a finite lattice, it can become important to include lattice corrections to the tensors $\Gamma$ \cite{Cucchieri:2006tf,Cucchieri:2004sq}. This prescription is used in the following for all the vertices.

The ghost-$W$ vertex is the expectation value
\be
G_\mu^{c\bar cW\; abc}(p,q,k)=\langle c^a(p)\bar c^b(q) W_\mu^c(k)\rangle\nn.
\ee
\no For the SU(2) gauge group there is only the tree-level color structure. In Landau gauge, furthermore, only the tensor component transverse in the gluon momentum is accessible. Hence, there is a single dressing function. It is projected out by choosing for $\Gamma$ the lattice version of the tree-level tensor, see \cite{Cucchieri:2006tf}.

For the three-$W$ vertex, 
\be
G_{\mu\nu\rho}^{WWW\; abc}(p,q,k)=\langle W_\mu^a(p) W_\nu^b(q) W_\rho^c(k)\rangle\nn
\ee
\no the situation is more complicated, as there are four independent transverse tensor structures \cite{Ball:1980ax}. Here only the tensor component of the tree-level tensor structure will be used, again including lattice corrections \cite{Cucchieri:2006tf}. As stated in \cite{Cucchieri:2006tf}, it is this tensor structure which is the most relevant one in most contemporary studies using functional methods.

Finally, the Higgs-$W$ vertex
\be
G_\mu^{\phi^\dagger\phi W\; ija}(p,q,k)=\langle \phi_i(p)\phi^\dagger_j(q)W_\mu^a(k)\rangle\nn
\ee
\no has a number of peculiarities, which require attention. It is, in principle, as simple as the ghost-$W$ vertex, since there is only one independent tensor structure transverse to the gluon momentum contributing. However, in the denoted form, it is not a flavor-invariant. As the corresponding symmetry is unbroken, the expectation value vanishes. To circumvent this problem, a flavor-invariant expectation value must be used, given by
\be
G_\mu^{\phi^\dagger\phi W\; ija}(p,q,k)=\langle X_{ik}(p) X_{kj}^\dagger(q) W_\mu^a(k) \rangle\nn,
\ee
\no based on the prescription \pref{x}. This vertex can, up to a normalization, still be projected in the same way as before, i.\ e.\ with a differently normalized tree-level vertex, to obtain the tensor structure. The corresponding tree-level tensor, including lattice corrections, is
\be
\Gamma^{\tl\; ija}_\mu(p,q,k)=\frac{iga}{6}\tau^a_{ij}\sin\frac{\pi}{N}(P-Q)_\mu\cos\frac{\pi}{N}(P+Q)_\mu\nn,
\ee
\no where $P$ and $Q$ are the integer-valued lattice momenta. This completes the list of vertex dressing functions to be calculated.

For three-point functions there are three independent kinematic variables. These will be chosen here to be the magnitude of the $W$ momentum and the particle momentum in the vertices. For the three-$W$ vertex, due to Bose symmetry, the choice is arbitrary. The third parameter is then the angle between these two momenta. Given the available resources, it was not possible to calculate all the possible choices. Thus, here only two particular important kinematical configurations will be discussed, the symmetric one and the orthogonal one \cite{Cucchieri:2006tf}.

The symmetric one is at an angle of $\pi/3$, and thus all three momenta have equal size. This is the configuration usually employed to derive running couplings from the three-point functions.

The second has an angle of $\pi/2$, and thus the two selected momenta are orthogonal to each other. This is the configuration with the largest integration measure in loop integrals, and should therefore give an idea about the dominating contribution from this vertex.

Note that all vertices renormalize multiplicatively.

Unfortunately, even the three-point vertices require, depending on the bare parameters and the types of the involved fields, one to two orders of magnitude more statistics than the propagators to achieve the same level of statistical error. It was hence not possible to investigate the vertices for all set of lattice parameters where the propagators have been studied, but only three different examples have been chosen.

\subsection{Bound states}\label{s:bs}

A detailed discussion of the bound states will be given elsewhere \cite{Maas:unpublished3}. However, to classify the dominant physics aspects in section \ref{sphase}, as well as to set the scale in section \ref{s:configs}, it is necessary to obtain some bound state information, most notably the masses of the $0^+$ flavor singlet and $1^-$ flavor triplet ground states. For the sake of completeness, here the procedure to determine them, essentially the one of \cite{Maas:2012tj} extending the one of \cite{Langguth:1985dr,Jersak:1985nf,Philipsen:1996af}, will be detailed. Note that only a rather rough determination of these masses is necessary for most purposes of the present work, and hence, e.\ g., the error on the lattice spacing will be suppressed throughout, since it is always at the few percent level or less.

In the $0^{+}$ channel several energy levels rather close by are found. To disentangle them, a basic variational analysis is performed \cite{Gattringer:2010zz}, using just two operators. One is the Higgsonium operator
\be
{\cal O}^{0^+}(x)=\phi_i^\dagger(x)\phi^i(x)=\rho(x)\label{higgsonium},
\ee
\no the other the $0^{+}$ $W$-ball state created by the plaquette \pref{plaq} as the lattice discretization of $W_\mn^a W_\mn^a$ \cite{DeGrand:2006zz}.

Since all bound state operators are very noisy, they have been four times APE smeared, i.\ e.\ the operators have been measured using the smeared links and Higgs fields \cite{Philipsen:1996af}
\bea
U_\mu(x)^{(n)}&=&\frac{1}{\sqrt{\det R_\mu(x)^{(n)}}}R_\mu(x)^{(n)}\nn\\
R_\mu^{(n)}&=&\alpha U_\mu(x)^{(n-1)}+\frac{1-\alpha}{2(d-1)}\nn\\
&&\times\sum_{\nu\neq\mu}\left(U^{(n-1)}_\nu(x+e_\mu)U^{(n-1)+}_\mu(x+e_\nu)U^{(n-1)+}_\nu(x)\right.\nn\\
&&\left.+U_\nu^{(n-1)+}(x+e_\mu-e_\nu)U_\mu^{(n-1)+}(x-e_\nu)U_\nu^{(n-1)}(x-e_\nu)\right)\nn\\
\phi^{(n)}&=&\frac{1}{1+2(d-1)}\left(\phi^{(n-1)}\right.\nn\\
&&+\left.\sum_\mu(U^{(n-1)}_\mu(x)\phi^{(n-1)}(x+e_\mu)+U^{(n-1)}_\mu(x-e_\mu)\phi^{(n-1)}(x-e_\mu))\right)\nn,
\eea
\no with $\alpha=0.55$ and $d=4$ and four iterations $n=4$.

To disentangle the ground state and the first excited state the correlation matrix of the two most-smeared operators of both types has been used to determine the eigenvalues and, as a cross-check, the eigenvectors. The lighter mass, obtained from a fit of type
\be
C(t)=A\cosh\left(am\left(t-\frac{N}{2}\right)\right)+B\cosh\left(an\left(t-\frac{N}{2}\right)\right)\label{stdfit},
\ee
\no has then been assigned to the ground state.

The usually more cleaner vector state was obtained using the operator
\bea
&&{\cal O}^{1^-}_{a\mu}=V_\mu^a,\label{wl}\\
&&=\tr \tau_a \det(-X(x))^\alpha X^\dagger(x) \exp(i\tau_bW^b_\mu(x))\det(-X(x+e_\mu))^\alpha X(x+e_\mu)\nn
\eea
\no again with the four times smeared operators. Note that the index $a$ is a flavor index. The power $\alpha$ is arbitrary, and does not change the quantum numbers, but changes the influence of excited states and statistical noise \cite{Jersak:1985nf}. Here, $\alpha=-1/2$ has been used, which suppresses excited states to some extent, but not as much as $\alpha=-1$, which makes the operator only dependent on $\varphi$. This choice was mainly made for the sake of the investigations to be discussed in \cite{Maas:unpublished3}. A fit to identify the ground state has then been performed using again \pref{stdfit}.

\section{QCD-like vs.\ Higgs-like}\label{sphase}

As noted already, the (lattice) theory has a continuously connected phase diagram \cite{Fradkin:1978dv}. Thus, though there might be exponentially large quantitative changes, the qualitative physics is the same throughout the phase diagram. Especially, there is no distinction of a Higgs phase and a confinement phase, signaled by the Higgs expectation value, as in the classical case. This is most easily seen in the non-aligned gauge used here, as there the Higgs expectation value is always zero, while it changes in an aligned gauge. But even the position of change in a fixed gauge is not unique, as it depends on the local part of the gauge \cite{Caudy:2007sf}. Nonetheless, there is a phase transition in the phase diagram, but it ends at a critical end-point, and therefore does not separate phases \cite{Langguth:1985eu,Evertz:1985fc,Bonati:2009pf}.

However, there exist two regions of the phase diagram, in which the physics shows quantitatively a distinctively different behavior. The most marked difference is the ordering of the ground states of the $0^+$ and $1^-$ channels, which changes between them \cite{Langguth:1985eu,Evertz:1985fc,Maas:unpublished3}. Especially, deep in the regime where in most aligned gauges the Higgs expectation value does not vanish\footnote{In the non-aligned gauge used here, it is, of course, always zero. Instead, an equivalent observable is the relative alignment $\langle\int dx\phi(x)^\dagger\int dy\phi(y)\rangle$ \cite{Caudy:2007sf}. A vanishing of this quantity in the infinite-volume limit corresponds to a vanishing Higgs expectation value in the corresponding aligned gauge, and vice versa.} the $1^-$ state is lighter, while in the other domain the $0^+$ state is lighter. In the cross-over and phase transition region, where also in some gauges there is always a phase transition and in some not, the two masses are (nearly) degenerate. Furthermore, in the domain where the $0^+$ state is lighter, a non-negligible intermediate distance string tension can be observed, before string-breaking sets in \cite{Knechtli:1998gf,Knechtli:1999qe}.

This will therefore be used here to define operationally a QCD-like domain (QLD) and a Higgs-like domain (HLD), away from the cross-over region (COR), by the decision whether $m_{1^-}/m_{0^+}$ is larger than one, smaller than one, or approximately one, respectively. As so far the mass of the lighter state is always the lightest one in all the investigated channels in both domains \cite{Maas:unpublished3}, this lighter mass will be taken to define the scale.

To set the scale, as was noted in section \ref{s:configs}, requires a number of further considerations \cite{Maas:2013eh}. The aim will be to obtain scales which are familiar from the electroweak physics. In this phase, there is a relation between the gauge-invariant $0^+$ and $1^-$ state's masses with the masses of the gauge-dependent Higgs and $W$ particles, in an expansion in the quantum fluctuations of the Higgs \cite{Frohlich:1980gj,Frohlich:1981yi}, which was confirmed on the lattice \cite{Maas:2012tj}, and which will be again confirmed more systematically here, at least for ratios of $m_{1^-}/m_{0^+}$ not too small compared to one.

These relations are valid only in an aligned gauge. Taking then the correlators \pref{higgsonium} and \pref{wl} in the continuum and expanding the Higgs field around its expectation value $vn^i$ $\phi^i(x)=\eta^i(x)+n^iv$, with $n^i$ some constant isospinvector, yields
\bea
&&\langle\phi_i^\dagger(x)\phi^i(x)\phi_j^\dagger(y)\phi^j(y)\rangle\nn\\
&\approx& v^4+4v^2(c+\langle\eta^\dagger_i(x) n^i n_j^\dagger\eta^j(y)\rangle)+{\cal O}(\eta^3)\label{correl},
\eea
\no and\footnote{Note that this result is independent of the power $\alpha$ in \pref{wl}, as to this order the determinant is just a constant, proportional to $v^2$.}
\bea
&&\langle(\tau^a\varphi^\dagger D_\mu\varphi)(x)(\tau^a\varphi^\dagger D_\mu\varphi)(y)\rangle\nn\\
&\approx& \tilde{c}\tr(\tau^a\tilde{n}\tau^b\tilde{n}\tau^a\tilde{n}\tau^c\tilde{n})\langle W^b_\mu W^c_\mu\rangle+{\cal O}(\eta W)\label{correl2},
\eea
\no with $c$ and $\tilde{c}$ are some constants, and $\tilde{n}$ the SU(2) matrix corresponding to $n$. Thus, up to this order, the masses, defined by the poles of the correlators, on both sides have to coincide. Hence, in the domain relevant to the standard model, a description in terms of the gauge-invariant and gauge-dependent degrees of freedom give an equally good picture of the physics, explaining the great success of perturbation theory. As will be seen below, this relation does not hold throughout the phase diagram. Nonetheless, this will be used to motivate setting in the HLD the scale such that the $1^-$ ground state mass will be 80.375 GeV. To obtain comparable scales also in the QLD, the scale will be set there by setting the ground state mass of the $0^+$ to the same value.

The only remaining problem is now that there are three independent parameters in the theory, and in principle a third external input is necessary. At the current time, no quantity is both experimentally and theoretical in lattice terms  good enough under control to serve as this input parameter. However, due to the absence of QED already the $W$-$Z$ mass splitting is missing. Moreover, the running gauge coupling, as will be seen below, runs much faster in the present theory as in the standard model, due to the lack of fermions and therefore a much larger $\beta_0$ \cite{Maas:2013eh}. Hence a quantitative comparison to the standard model is at the current time anyhow only of limited reliability, a problem recognized also in other investigations \cite{Wurtz:2013ova}.

However, since we are interested here in understanding the theory as such, and not yet too much the experimental situation, we will not constraint us to a single line of constant physics (LCP), but rather will use a large set of different points throughout the phase diagram, to understand the behavior. As will be shown below, it turns out that most of the properties of the propagators and three-point vertices are actually mainly controlled by the ratio $m_{1^-}/m_{0^+}$, and therefore fixing the third parameter uniquely seems anyhow to be of little relevance, at least in the part of the phase diagram investigated here.

\begin{figure}
\centering
\includegraphics[width=0.5\linewidth]{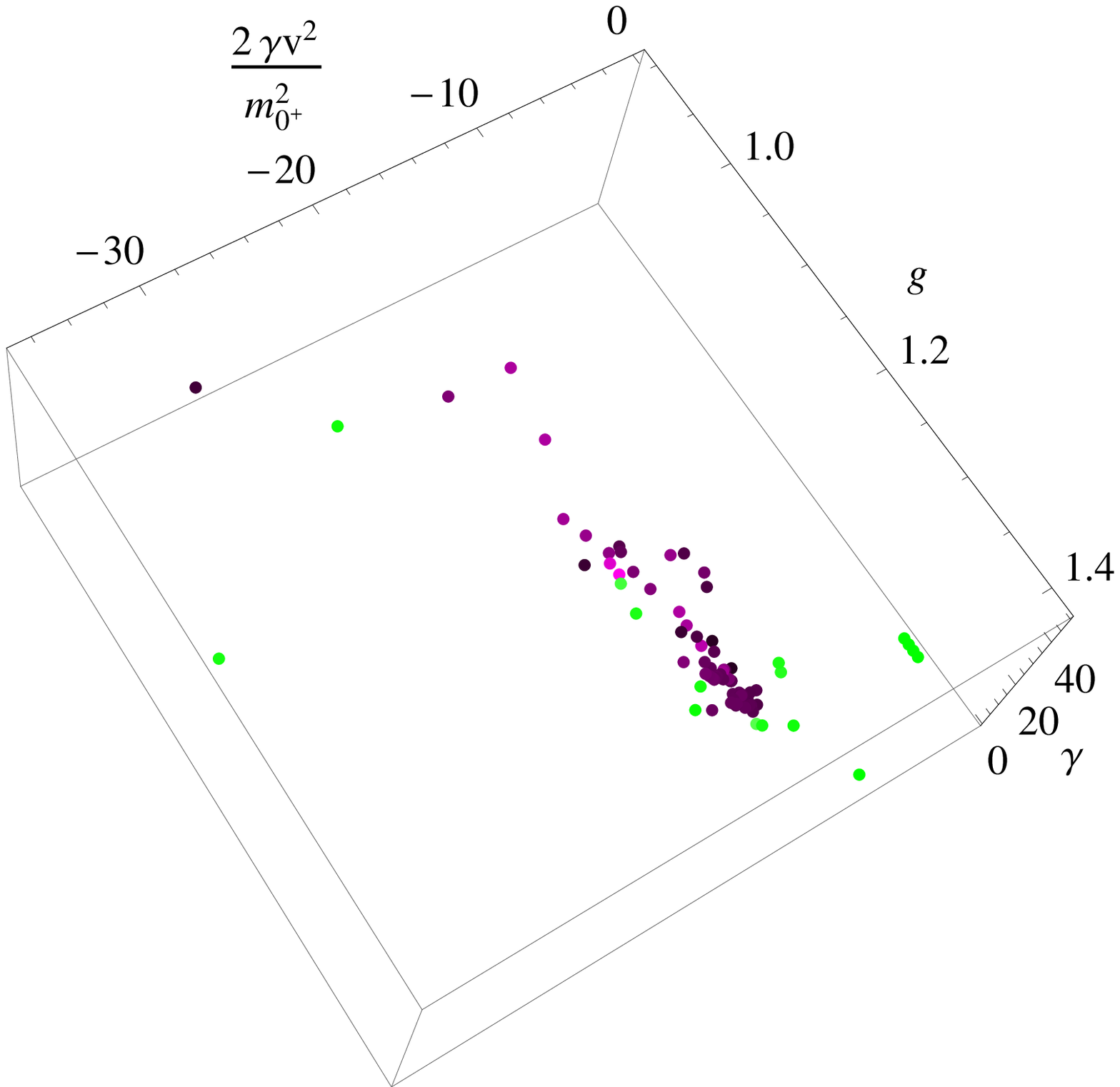}\includegraphics[width=0.5\linewidth]{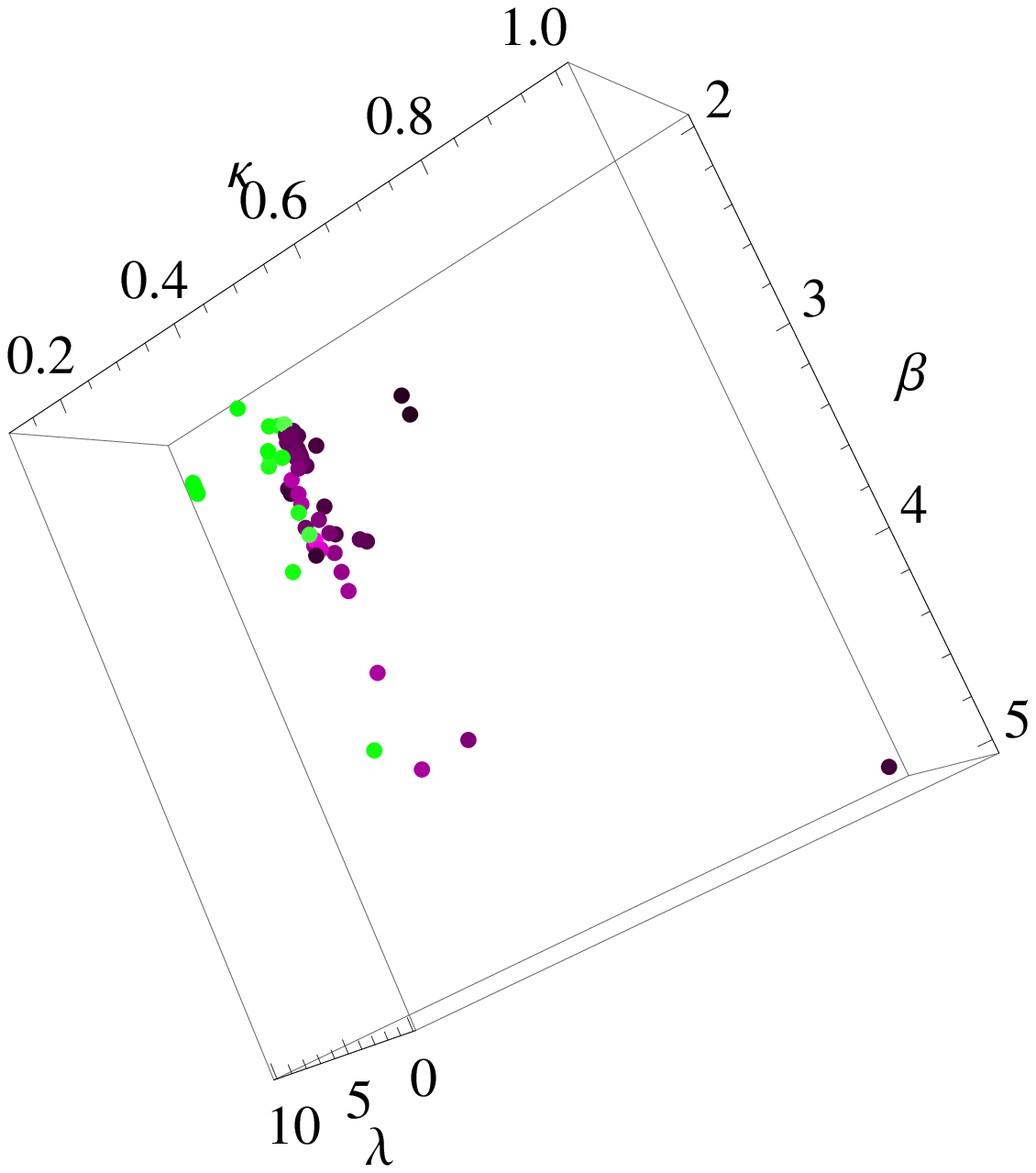}
\caption{\label{lcp}The phase diagram of the Yang-Mills-Higgs theory as a function of bare gauge coupling, Higgs 4-point coupling, and the bare Higgs mass in units of the $0^+$ mass. Green points are confinement-like, and purple points are Higgs-like. The lighter the points, the smaller is the lattice spacing. The right-hand plot shows the same in terms of the lattice bare parameters of inverse gauge coupling, hopping parameter, and four-Higgs coupling, see \pref{action} for their relation.}
\end{figure}

This part of the parameter region is shown in figure \ref{lcp}. It is visible, how the phase diagram disconnects into the two parts, the HLD and the QLD. Interestingly enough, but not surprising due to the additive mass renormalization, the QLD region persists even deep into the negative $m_0^2$ region, where classically already the Higgs effect would be operative.

\begin{table}
\caption{\label{configs}The lattice parameters $\beta$, $\kappa$, and $\lambda$ for the employed configurations, together with the masses of the $1^-$ and $0^+$ ground states, the derived ratio and lattice spacing, and the classification. Various lattice volumes $N^4$ have been used, and the sizes are indicated in the corresponding figures. In addition, also the plaquette expectation value $\langle P\rangle$ and the value for the Higgs length $\langle\phi^\dagger\phi\rangle^\frac{1}{2}$ are displayed. Both quantities are unrenormalized and from $24^4$ lattices. Note that, because of the gauge interaction, a vanishing $\lambda$ did not create, even for a negative quadratic term, a runaway condition.}
\vspace{1mm}
\begin{tabular}{|c|c|c|c|c|c|c|c|c|}
\hline
\hline
Type & $\beta$ & $\kappa$ & $\lambda$ & $m_{0^+}$ & $m_{1^-}$ & $a^{-1}$ [GeV] & $\langle P\rangle$ & $\langle\phi^\dagger\phi\rangle$ \cr
\hline
QLD  & 2.3095  & 0.2668   & 0.5254    & 0.45(5)   & 1.08(1)   & 179            & 0.616787(3)  & 1.17838(1) \cr
\hline
QLD  & 2.221   & 0.125    & 0         & 1.44(2)   & 3.3(3)    & 56             & 0.577412(21) & 1.44969(1)  \cr
\hline
QLD  & 2.2171  & 0.3182   & 1.046     & 0.51(5)   & 0.57(1)   & 142            & 0.600879(4)  & 1.148630(6) \cr
\hline
\hline
HLD  & 2.2667  & 0.3141   & 1.043     & 0.68(5)   & 0.51(1)   & 158            & 0.614414(2)  & 1.147419(3) \cr
\hline
HLD  & 2.4728  & 0.2939   & 1.036     & 0.41(1)   & 0.296(4)  & 272            & 0.652354(5)  & 1.11696(1) \cr
\hline
HLD  & 2.3     & 0.31     & 1.0       & 0.74(1)   & 0.48(1)   & 168            & 0.6228101(8) & 1.152029(2) \cr
\hline
HLD  & 2.3634  & 0.3223   & 1.066     & 1.12(7)   & 0.53(1)   & 153            & 0.642131(2)  & 1.171088(4) \cr
\hline
HLD  & 2.3     & 0.32     & 1.0       & 1.04(2)   & 0.548(3)  & 148            & 0.632379(1)  & 1.206336(2) \cr
\hline
HLD  & 2.8     & 0.318    & 1.2       & 1.21(1)   & 0.414(3)  & 194            & 0.707624(2)  & 1.154240(4) \cr
\hline
HLD  & 2.7984  & 0.2954   & 1.317     & 0.47(3)   & 0.219(2)  & 368            & 0.701833(2)  & 1.09106(1) \cr
\hline
HLD  & 2.3579  & 0.3208   & 1.010     & 1.2(1)    & 0.26(5)   & 308            & 0.641783(2)  & 1.18108(1) \cr
\hline
HLD  & 2.3827  & 0.3176   & 1.018     & 1.10(1)   & 0.33(7)   & 244            & 0.645325(5)  & 1.173572(9) \cr
\hline
\hline
\end{tabular}
\end{table}

In the following now the propagators and 3-point vertices will be studied for a subset of the displayed systems. This subset is listed in table \ref{configs}. However, because of the statistics required, it was not possible to investigate for all settings in addition also the 3-point vertices. Hence, as a representative selection, three settings have been chosen. These correspond to a system deep inside the QLD, one with the physical $m_{1^-}/m_{0^+}$ ratio of roughly 0.64, and one with a small ratio of 1/3, corresponding to a Higgs mass of 243 GeV. Though not yet in the range where the Higgs self-interaction is very strong, such systems have not been included here, the Higgs can decay into two on-shell $W$, opening new decay channels. It is an interesting question, whether this manifests itself in the three-point functions.

As it is not entirely trivial to follow the LCPs, due to the fine-tuning problem especially in $\kappa$, at the current time only a very limited amount of different lattice spacing effects can be studied.

\section{Propagators}\label{sprop}

\subsection{Gauge boson}

\begin{figure}
\centering
\includegraphics[width=\linewidth]{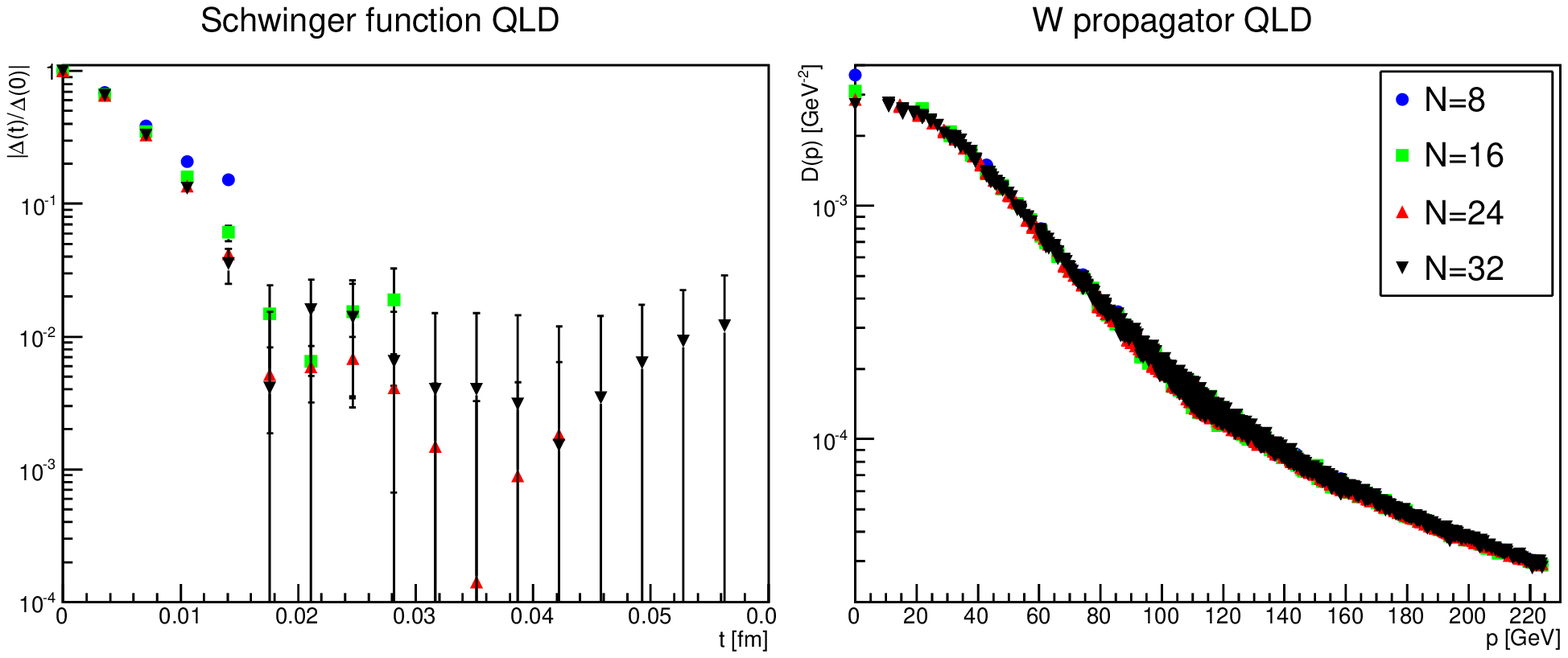}
\includegraphics[width=\linewidth]{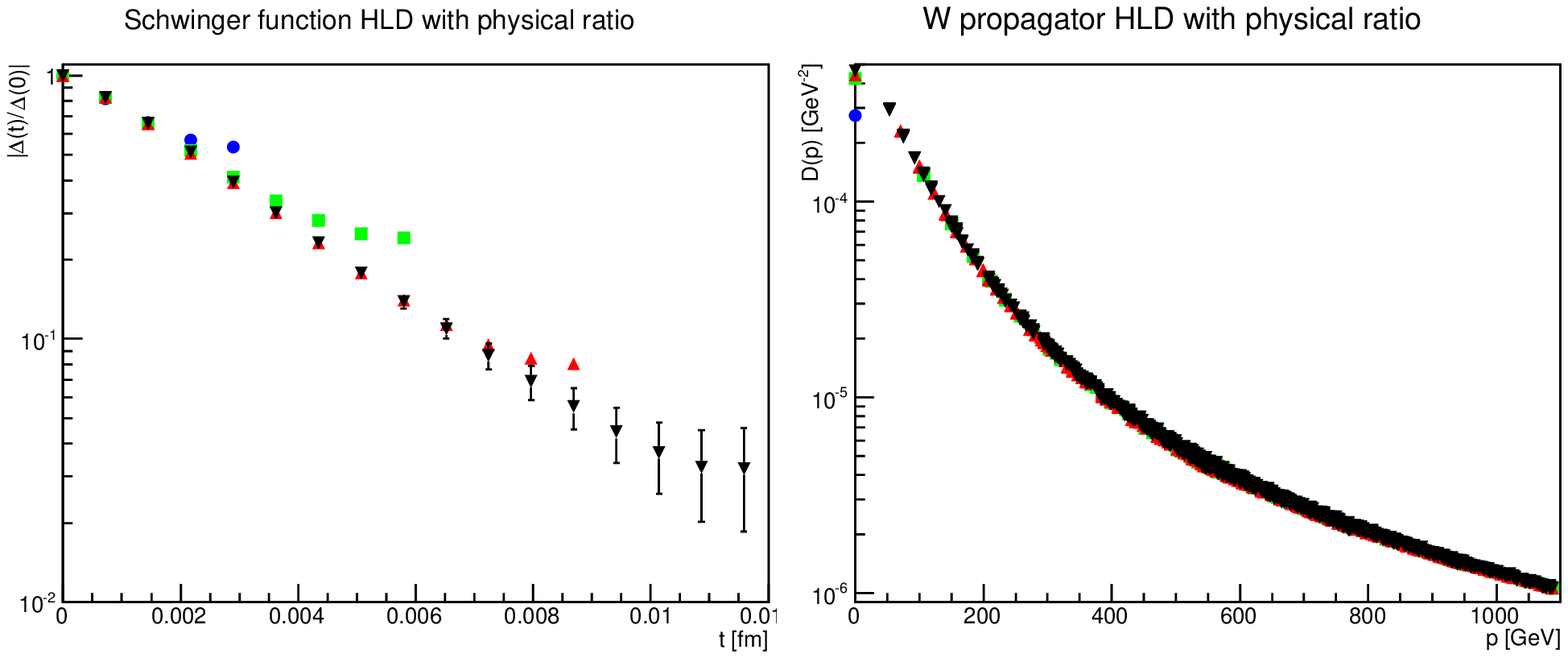}
\includegraphics[width=\linewidth]{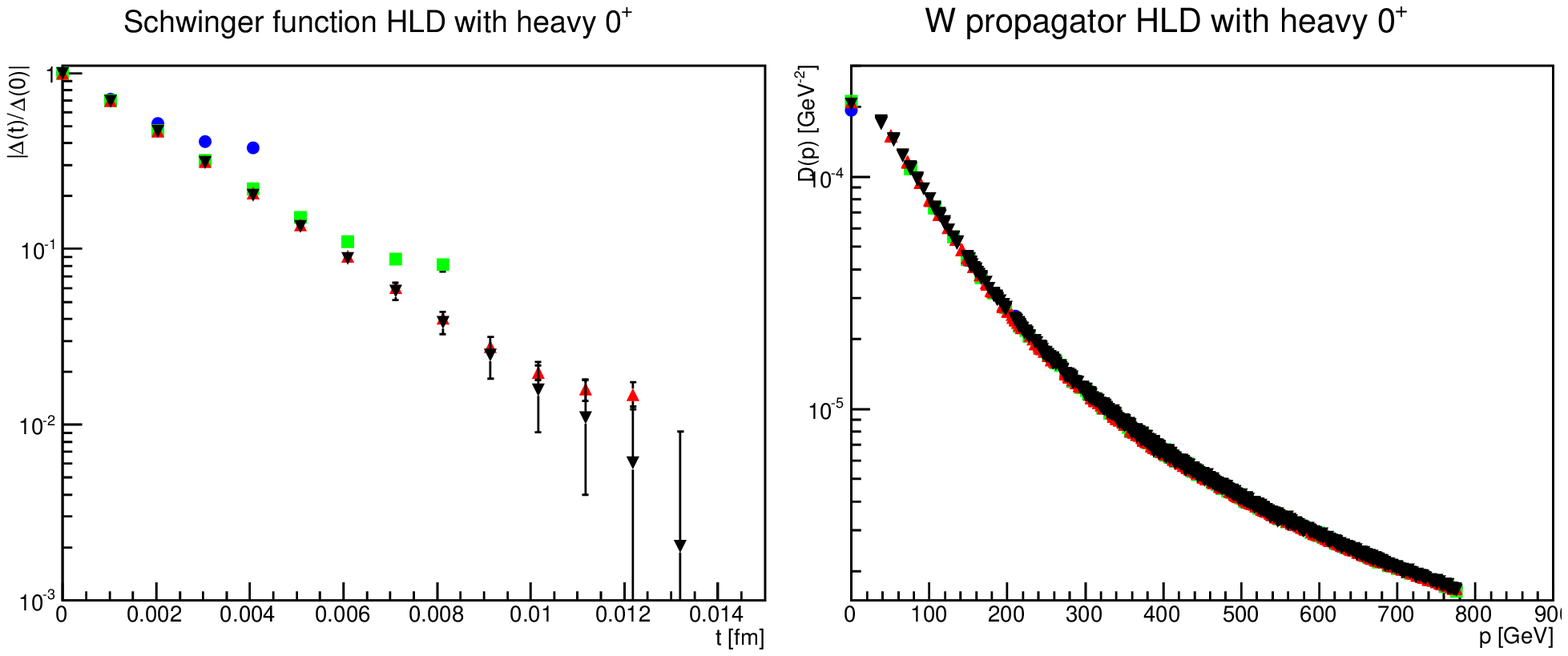}
\caption{\label{fig:wv}The gauge boson propagator in position space (left panel) and momentum space (right panel) for different volumes. The top panel is in the QLD with $m_{1^-}/m_{0^+}=2.2$, the middle panel has the physical mass ratio $m_{1^-}/m_{0^+}=0.72$, and the bottom panel is for a large Higgs mass $m_{1^-}/m_{0^+}=0.31$, both in the HLD.}
\end{figure}

The simplest possible object, which can be studied, is the gauge boson, i.\ e.\ $W$ propagator. Since in the Yang-Mills case it is severely affected by finite-volume effects \cite{Maas:2011se}, first lattice artifacts will be studied. These volume-effects are shown in figure \ref{fig:wv}. First of all, it is visible that the  finite-volume effects in the HLD and QLD have opposite effects, i.\ e.\ the propagator is suppressed the larger the volume in the QLD and enhanced the larger the volume in the HLD. Furthermore, the finite-volume effects in the HLD diminish with increasing $0^+$ mass.

The behavior in the QLD is quite similar to the one observed in Yang-Mills theory. The one in the HLD is fundamentally different\footnote{At very small volumes, the same effect is also observed in the Yang-Mills case \cite{Mandula:1987rh,Maas:2011se,Maas:unpublished}. However, given the masses in lattice units of the lightest physical state in the HLD calculations here, the volumes cannot be considered so small.}, but they appear to converge quicker than in the QLD case. In any case, the value of the $W$ propagator at zero momentum is to be considered unreliable, though its volume-dependence itself maybe of interest in principle \cite{Fischer:2007pf}.

Of course, at large times the position-space correlator shows the typical deviations for a finite volume in all cases.

Note that while only a selection of lattice parameters are shown here, at least the finite volume behavior and, where available, the lattice spacing effects have been investigated for many more of the systems shown in figure \ref{lcp}. In no case a qualitative different pattern has been observed. This statement holds also true for all the results on the propagators to be studied below, and will therefore not be repeated again.

\begin{figure}
\centering
\includegraphics[width=\linewidth]{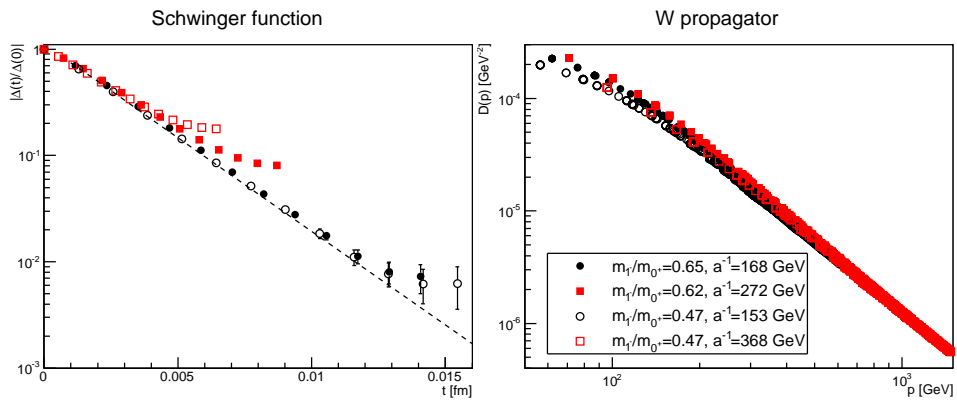}
\caption{\label{fig:wa}The gauge boson propagator in position space (left panel) and momentum space (right panel) for different mass ratios $m_{1^-}/m_{0^+}$ on 24$^4$ lattices and for two different lattice spacings.}
\end{figure}

Considering the dependence on the lattice spacing is more complicated, as it is not entirely trivial to be sure to be on the same LCP, mainly due to the lack of a third observable, and the fact that other states are heavy and therefore their mass determination tends to be also affected by lattice artifacts \cite{Maas:unpublished3}. Comparing anyway two cases in the HLD with different lattice spacing but the same ratio $m_{1^-}/m_{0^+}$ in figure \ref{fig:wa} shows that nonetheless there is very little difference between the two sets of lattice parameters. This indicates that for the present purpose the influence of this type of lattice corrections is small, and that the third physical parameter plays not a too big role here. Of course, further systematic studies are required to make this statement more reliable. However, already in the Yang-Mills case \cite{Maas:2011se} lattice-spacing effects have been found to be at the few percent level for two-point and three-point correlation functions.

\begin{figure}
\centering
\includegraphics[width=\linewidth]{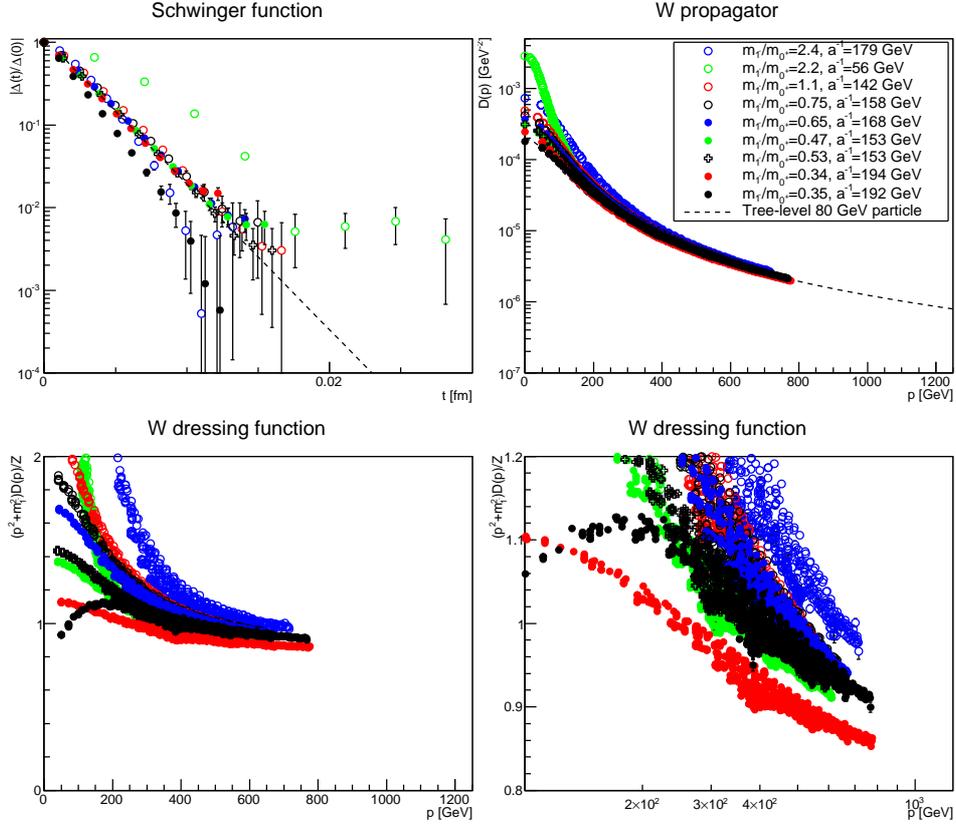}
\caption{\label{fig:w}The top panels show the gauge boson propagator in position space (left panel) and momentum space (right panel) for different mass ratios $m_{1^-}/m_{0^+}$ on 24$^4$ lattices. The tree-level result is for the infinite-volume case. The lower panels show the ratio to the expected dressing function, where $Z=1.4$ is a wave-function renormalization constant. Note that the propagators are unrenormalized.}
\end{figure}

Finally, the $W$ propagator for different values of the ratio $m_{1^-}/m_{0^+}$ is displayed in figure \ref{fig:w}. A number of very interesting observations are immediately possible. The first is that at large momenta all propagators tend to the same asymptotic behavior. This is expected, as the mass scale generated by the Higgs effect should become irrelevant at large energies. However, this common behavior is not that of a mass-less particle, but there are logarithmic corrections, which are particular visible in the lower-right panel. These stem partly from renormalization effects. The fact that also the QLD propagators join in the same behavior emphasize that the mass is not a hard mass, and it diminishes quicker at high energies than an ordinary mass function would do.

The second is that the behavior of the space-time-correlation functions is markedly different for the QLD and the HLD. While in the HLD it is positive, there is a zero-crossing observed in the QLD. The latter is also characteristic for Yang-Mills theory \cite{Maas:2011se,Fischer:2008uz}, as well as QCD \cite{Bowman:2007du,Alkofer:2003jj}. It implies positivity violation in the spectral density.

The result for the HLD for the space-time correlator is also somewhat surprising. While at small masses they all coincide with the behavior expected because of \pref{correl2}, i.\ e.\ they decay like a massive particles at long time with the mass $m_{1^-}$, this does not appear to be the case if the $0^+$ mass exceeds the $1^-$ by more than a factor of two.

\begin{figure}
\centering
\includegraphics[width=\linewidth]{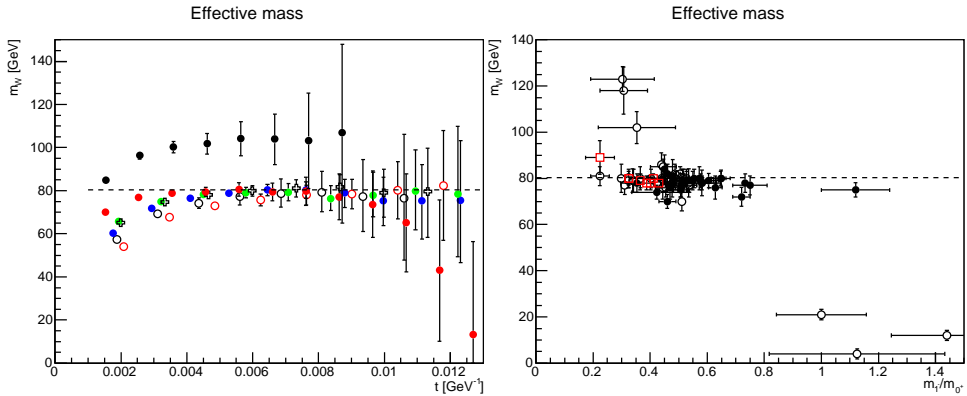}
\caption{\label{fig:wm}In the left-hand panel the effective masses for the propagators in the HLD phase of figure \ref{fig:w} are shown. Points with too large errors are suppressed. In the right-hand panel the masses obtained from the plateaus are shown for all lattice parameters in the HLD also shown in figure \ref{lcp}. In this case also the errors from the scale setting have been included. Full circles have $0^+$ masses in lattice units below 1, open circles between 1 and 3/2, and red squares above 3/2.}
\end{figure}

To make this statement more quantitative, the effective mass
\be
m(t)=-\ln\frac{\Delta(t)}{\Delta(t+a)}\nn,
\ee
\no is plotted in the left-hand panel of figure \ref{fig:wm} for the HLD case. The first observation is that there is a plateau, corresponding to a mass. But the approach to the plateau is from below, instead of above. This is not possible for a physical particle. However, the $W$ boson is also in the HLD gauge-dependent, and not subject to such constraints, like in the QLD. Physically, the origin of this phenomenon is that the mass of the $W$ is not a hard mass, but the propagator should vanish quicker than one with such a hard mass, transmuting into a massless particle at large momenta, i.\ e.\ short times. This was already visible in figure \ref{fig:w}. Hence, at short times a different decay is to be expected, and the transition leaves its mark in the effective mass behavior: The correlator shows a lighter instead of a heavier behavior at short distances.

At long times the behavior becomes massive. Extracting from the plateaus the effective mass yields the results shown in the right-hand panel of figure \ref{fig:wm}. In the transition region from the QLD to the HLD the relation \pref{correl2} is strongly violated. This is not surprising, as it does not hold in the QLD, where there is no pole mass in the conventional sense at all. In the interval $1>m_{1^-}/m_{0^+}>1/2$, i.\ e.\ between entering the HLD and while the $0^+$ remains stable against the decay in two $1^-$, the relation \pref{correl2} is fulfilled within errors. Starting at $m_{1^-}/m_{0^+}<1/2$, two branches are observed, one in which the relation \pref{correl2} remains fulfilled, and one where this is not the case. As the relation \pref{correl2} is the requirement that the observable $1^{-}$ state can be identified with the elementary $W$ boson, this implies that on the second branch a perturbative description is no longer reliable in the conventional sense. This would be at an unexpected small value of the mass of the $0^+$; usually this is scheduled in perturbation theory to occur at a mass scale of more than 750 GeV \cite{Bohm:2001yx}.

Investigating the lattice parameters show that the branch with a fulfilled relation \pref{correl2} has smaller bare lattice gauge couplings, while the other branch has larger ones. Note that this has no implications for the lattice spacing, and on both branches similar lattice spacings are observed. In fact, the bare lattice couplings bear no physical meaning, and it is required to investigate other quantities to understand the origin of this difference.

\subsection{Ghost}

One possibility to translate the bare coupling to a physical one is by determining the corresponding running (gauge) coupling. In Landau gauge, this is simplified in the here deployed miniMOM renormalization scheme \cite{vonSmekal:2009ae}, as it is possible to obtain it just from the ghost and the $W$ boson propagator. This entails to determine the ghost propagator, which will be done in this section, before assembling the full running coupling in the next section.

\begin{figure}
\centering
\includegraphics[width=\linewidth]{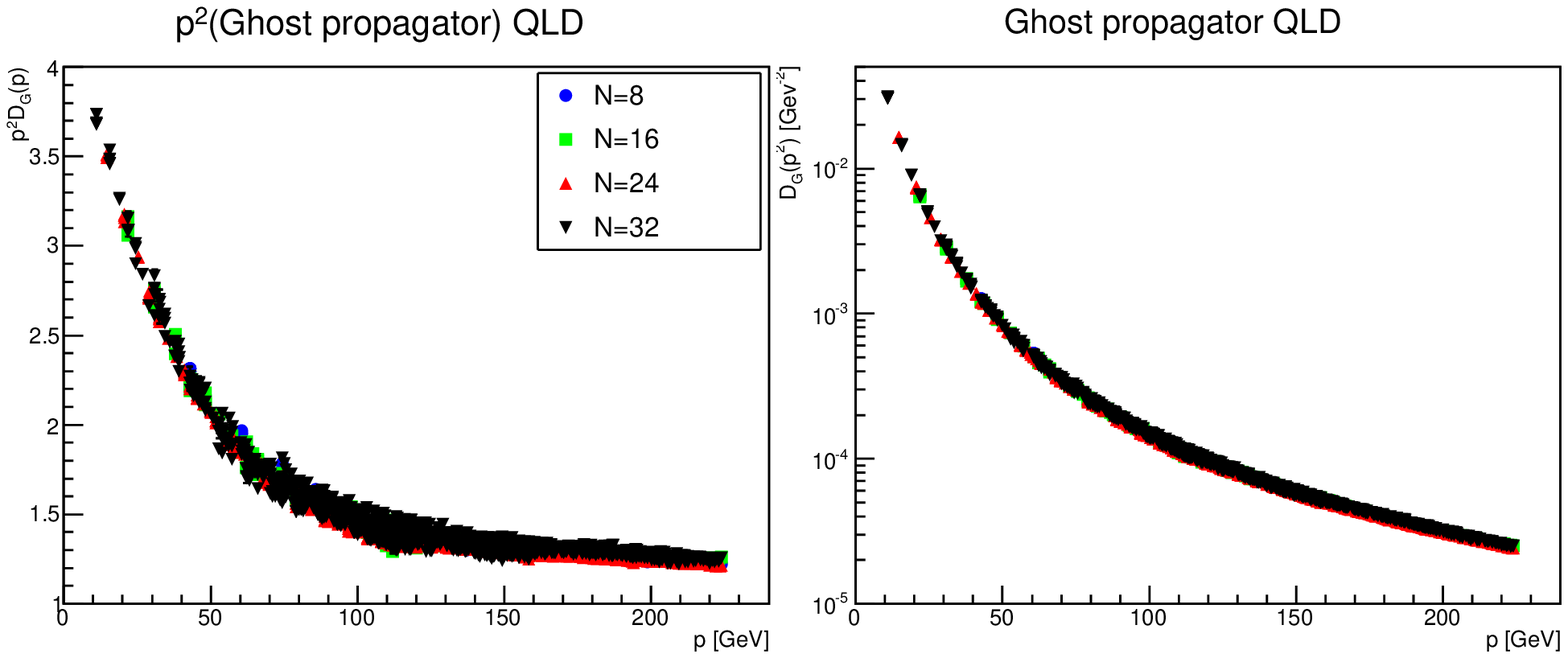}
\includegraphics[width=\linewidth]{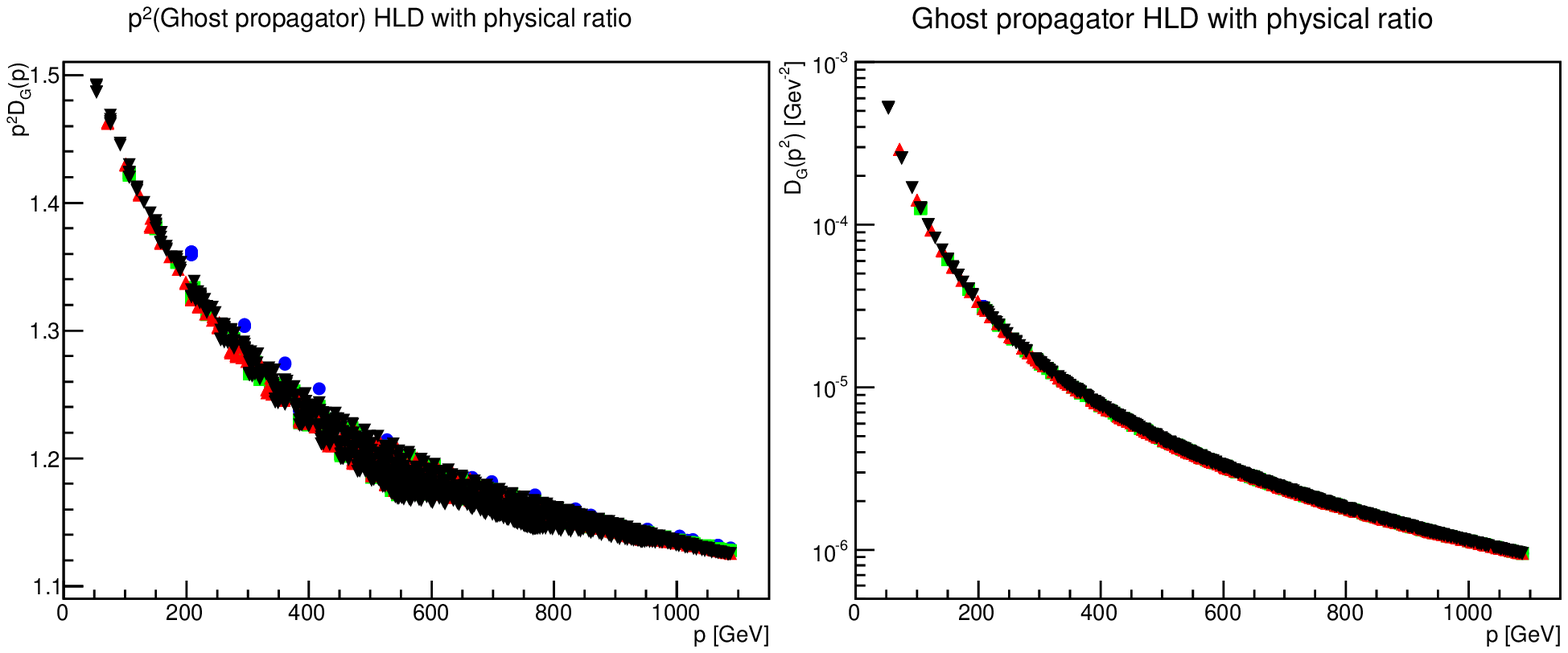}
\includegraphics[width=\linewidth]{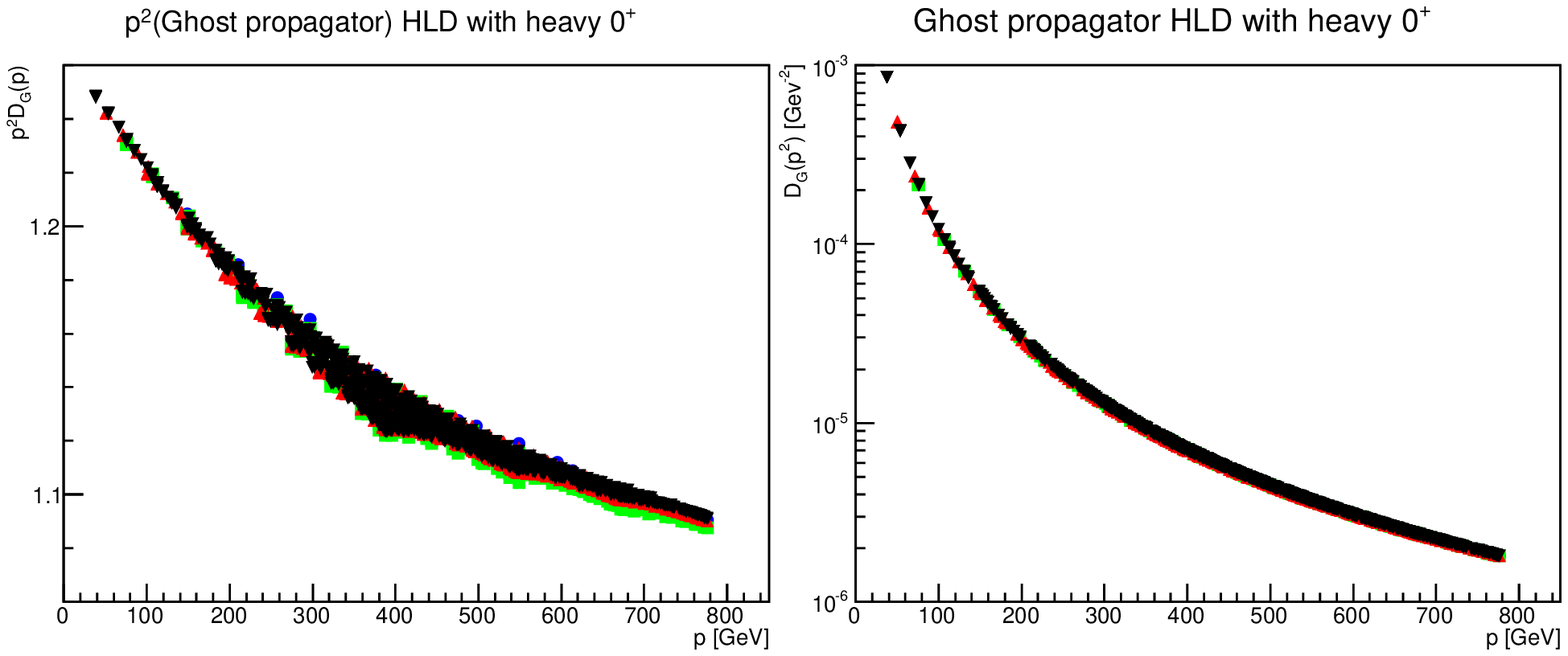}
\caption{\label{fig:gv}The ghost propagator (right panel) and dressing function (left panel) for different volumes. The top panel is in the QLD with $m_{1^-}/m_{0^+}=2.2$, the middle panel has the physical mass ratio $m_{1^-}/m_{0^+}=0.72$, and the bottom panel is for a large Higgs mass $m_{1^-}/m_{0^+}=0.31$, both in the HLD.}
\end{figure}

\begin{figure}
\centering
\includegraphics[width=\linewidth]{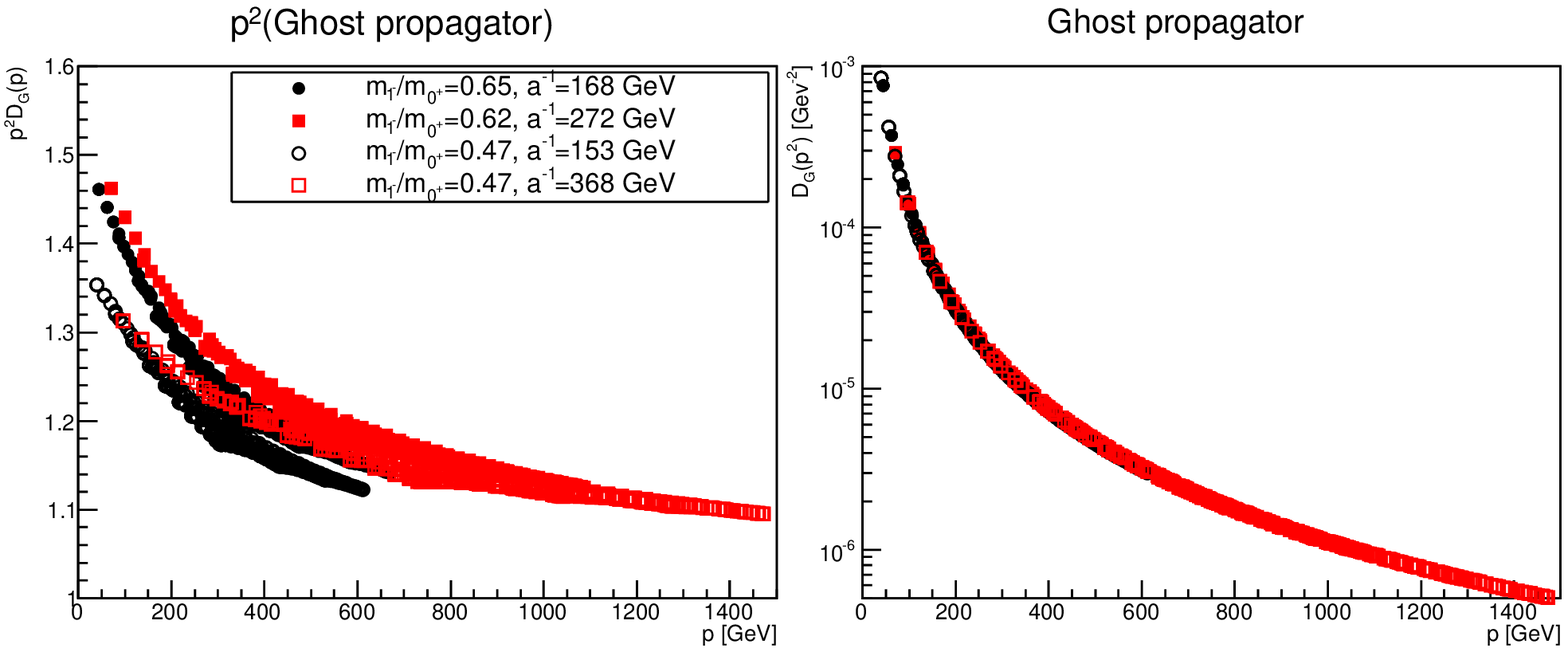}
\caption{\label{fig:ga}The ghost propagator (right panel) and dressing function (left panel) for different mass ratios $m_{1^-}/m_{0^+}$ on 24$^4$ lattices and for two different lattice spacings.}
\end{figure}

Once more, the experience with Yang-Mills theory warns to be wary of lattice artifacts. In the same manner as for the $W$ propagator, finite volume and lattice spacing effects are studied in figure \ref{fig:gv} and \ref{fig:ga}, respectively. It is visible that there is at most a slight volume dependence in all cases. However, the effect is similar to the one in Yang-Mills theory \cite{Maas:2011se}. There, despite an appearance as in the top panel of figure \ref{fig:gv}, the ghost propagator is found to be finite towards the infrared \cite{Cucchieri:2008fc,Bogolubsky:2009dc,Sternbeck:2007ug}, due to very slowly manifesting volume effects. It appears likely that the same is true here as well, at least in the QLD, given the similarities for the $W$ propagator. Of course, larger volumes would be necessary for a conclusive statement.

The situation is more pronounced in the lattice spacing case. The changes in lattice spacing displayed is not leading to more than a factor two in physical momenta. Nonetheless, the ghost propagator is substantially different from each other in this case, compared to the finite-volume effect. The reason for the somewhat stronger dependence is therefore not due to the change of volume. Furthermore, the behavior is non-monotonous in momentum, and thus cannot be cured by a multiplicative renormalization. It leads mainly to a weaker momentum-dependence towards larger momenta. The infrared region is less affected. Still, this a 10\% effect at most.

\begin{figure}
\centering
\includegraphics[width=\linewidth]{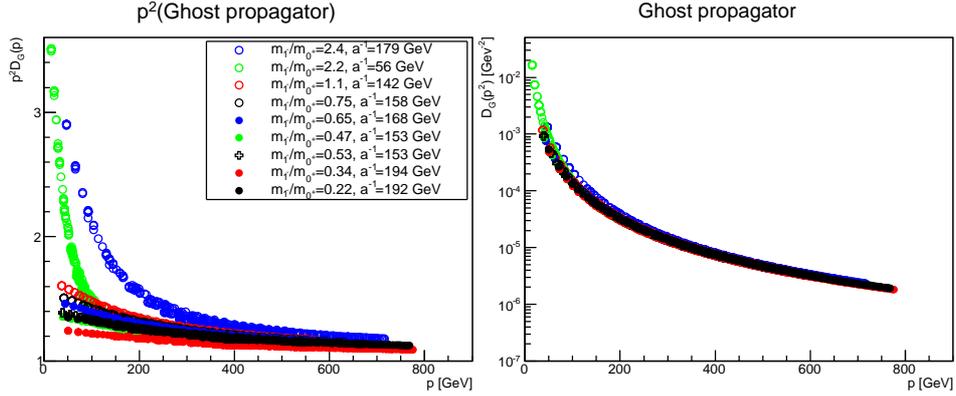}
\caption{\label{fig:ghp}The ghost propagator (right panel) and dressing function (left panel) for different mass ratios $m_{1^-}/m_{0^+}$ on 24$^4$ lattices. Note that the propagators are unrenormalized.}
\end{figure}

The ghost propagator is shown for different values $m_{1^-}/m_{0^+}$ in figure \ref{fig:ghp}. A drastic difference can be seen between the QLD and HLD. In the former case, the propagator shows a behavior resembling quite closely the one of Yang-Mills theory \cite{Maas:2011se}. As stated above, this makes it likely that it is infrared finite, as in the Yang-Mills case, though the volume-dependence is not yet conclusive.

The situation is quite different in the HLD, where the ghost propagator is much less infrared enhanced, and the deviation from a massless particle is extremely small. Such a masslessness is in agreement with perturbation theory in Landau gauge \cite{Bohm:2001yx}. It is also compatible with earlier indirect evidence based on the spectrum of the Faddeev-Popov operator \cite{Greensite:2004ke}, which was found to be close to the perturbative one. Finally, the remaining infrared enhancement seems to diminish with decreasing mass ratio $m_{1^-}/m_{0^+}$, and thus increasing Higgs mass. Note that the two branches observed for the $W$ propagator show no strongly distinct behavior for the ghost propagator.

\subsection{Running coupling}

Having both the ghost and the $W$ propagator at hand, it is possible to construct the running gauge coupling, which in the miniMOM scheme is given by \cite{vonSmekal:1997vx,vonSmekal:2009ae}
\be
\alpha(p^2)=\alpha(\mu^2) p^6 D_G(p^2,\mu^2)^2 D(p^2,\mu^2)\label{alpha},
\ee
\no and thus just entirely in terms of the propagators. The scale $\mu^2$ is the one where the (experimental) input value for the running coupling is selected. The PDG value \cite{pdg} is available at the $Z$ mass, however in a different scheme. Given that this is of the order of the involved masses, and the running coupling itself being dependent on the gauge, a direct translation is not feasible. Therefore, rather the ratio $\alpha(p^2)/\alpha(\mu^2)$ will be used here. Since the running coupling is just a product of the propagators, its lattice-artifact-dependence is just a  combination of the ones of the propagators, and it will therefore not be studied explicitly here.

\begin{figure}
\centering
\includegraphics[width=\linewidth]{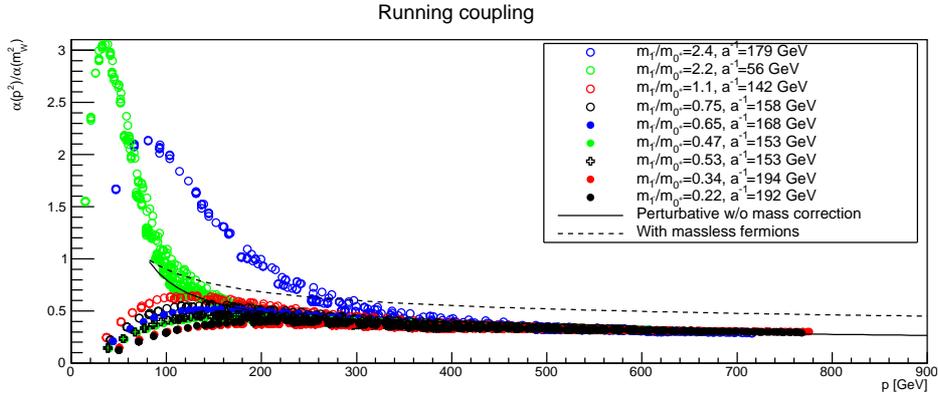}
\caption{\label{fig:alpha}The running coupling for different mass ratios $m_{1^-}/m_{0^+}$ on 24$^4$ lattices. Renormalization has been performed such as to agree with the perturbative running coupling at large momenta. See the text for details on the latter.}
\end{figure}

The resulting running coupling is shown in figure \ref{fig:alpha}. The first observation is that once more at large momenta all results agree very well with the leading-order massless running gauge coupling
\be
\frac{\alpha(p^2)}{\alpha(\mu^2)}=\frac{1}{1+\frac{1}{4\pi}\beta_0\ln\frac{p^2}{\mu^2}}\label{alphar},
\ee
\no where $\beta_0$ is the first coefficient of the $\beta$-function, which has a value of $43/6$ in the present theory. This coincides with the previous observation: At momenta large compared to the scale set by the Higgs mechanism, the behavior is the same for QLD and HLD, and essentially that of the massless theory. It should be noted that this behavior is markedly different from the also plotted case with the 12 species of standard model fermions included, for which $\beta_0$ has the value $19/6$. The ratio of both cases at 1.2 TeV is then still 0.578, which is larger than the ratio of the $\beta_0$s. Hence, in the full range the gauge coupling in the present theory runs faster than the one with fermions, and the gauge interactions would actually be stronger when including the fermions. Thus, the present theory has a substantially weaker integrated gauge interaction than the standard model, as already remarked in \cite{Maas:2013eh}.

Returning to the ultraviolet behavior, the far ultraviolet is rather universal. This is not surprising, as the propagators show in both the QLD and HLD the same behavior, despite their different analytic structure. Of course, if desired, the scheme could be redefined that in all cases the couplings would run to an infrared fixed point \cite{Fischer:2008uz,Aguilar:2008fh}, but this is rather cosmetic, and of no relevance here.

More interesting is the mid-momentum regime, i.\ e.\ momenta of the order of the bound-state masses between 50 and 250 GeV. Here there is a strong quantitative difference between the QLD and HLD. In the QLD the running coupling show a pronounced peak, signaling a large integrated strength, like in Yang-Mills theory \cite{Maas:2011se}. This integrated strength is the closest possible definition of the statement of a strongly interacting theory, as e.\ g.\ in QCD this integrated strength is responsible for chiral symmetry breaking \cite{Fischer:2006ub,Roberts:2000aa}. The situation is drastically different in the HLD. There, some maximum remains, though this is essentially by construction with an infrared and ultraviolet vanishing running coupling. The height of this maximum decreases continuously with the mass ratio $m_{1^-}/m_{0^+}$, and moves at the same time also to larger momenta. Thus, the integrated strength diminishes with decreasing ratio $m_{1^-}/m_{0^+}$. Note that this effect is independent of the branch at large Higgs mass: The integrated running coupling strength is not a monotonous function of the bare coupling. The latter therefore gives no indication of the interaction strength of the theory.

As a consequence, it would be expected that the gauge interaction becomes less relevant the smaller the ratio $m_{1^-}/m_{0^+}$ is. It remains to see whether this is true.

Note that there is no three-Higgs vertex in a non-aligned gauge, and there is, to our knowledge, no simple relation like \pref{alpha} for the four-Higgs interaction, so that no such calculation can be done for this running coupling. As stated already, a direct calculation is obstructed by the statistical noise.

\subsection{Higgs}

The last propagator is the Higgs propagator. As noted already in section \ref{s:propagators}, it is different from the $W$ and the ghost propagator in so far as that it requires also an additive mass renormalization. Due to the lack of extensive LCPs, it is not yet possible to study the renormalization properties in detail. This is possible in the quenched case and this will be discussed elsewhere \cite{Maas:unpublished}, though the upshot is that the renormalization is essentially what is expected from a perturbative calculation \cite{Bohm:2001yx}.

As a consequence, however, the masses extracted from the Higgs propagator space-time correlator depend on the renormalization scheme \prefr{scheme1}{scheme2} \cite{Maas:2012tj}. This problem did not surface in the relation \pref{correl} as to lowest order in the quantum fluctuations renormalization effects do not play a role. However, in the present lattice calculations all such quantum effects are included, and therefore checking \pref{correl}, in contrast to the case of the $W$ boson where no mass renormalization is necessary, is meaningless.
 
Of course, in a pole scheme this could be superficially cured by enforcing that the mass of the Higgs becomes the one of the $0^+$ in a kind of mimicking the pole/on-shell scheme of perturbation theory \cite{Ghinculov:1996py,Bohm:2001yx}. Then the mass is uniquely fixed by an observable. However, in a sense this is cheating, as this choice is arbitrary. This will nonetheless be made\footnote{Note that the situation could actually be worse, as the Nielsen identities ensuring gauge-invariance of the Higgs and $W$ masses in certain classes of gauges are actually not guaranteed to hold between different classes of gauges \cite{Nielsen:1975fs}, and the situation for non-aligned, and therefore genuinely non-perturbative \cite{Maas:2012ct,Lee:1974zg}, gauges is not yet settled.}.

\begin{figure}
\centering
\includegraphics[width=\linewidth]{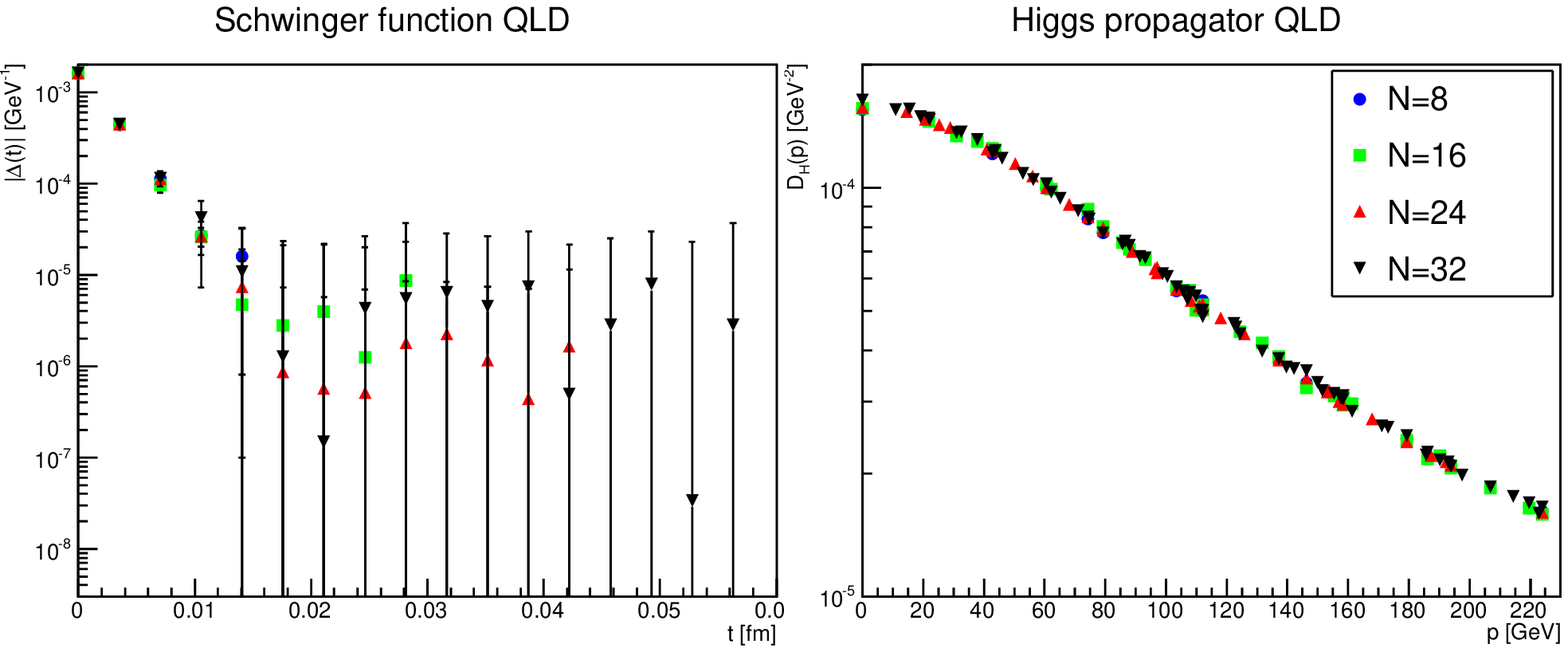}
\includegraphics[width=\linewidth]{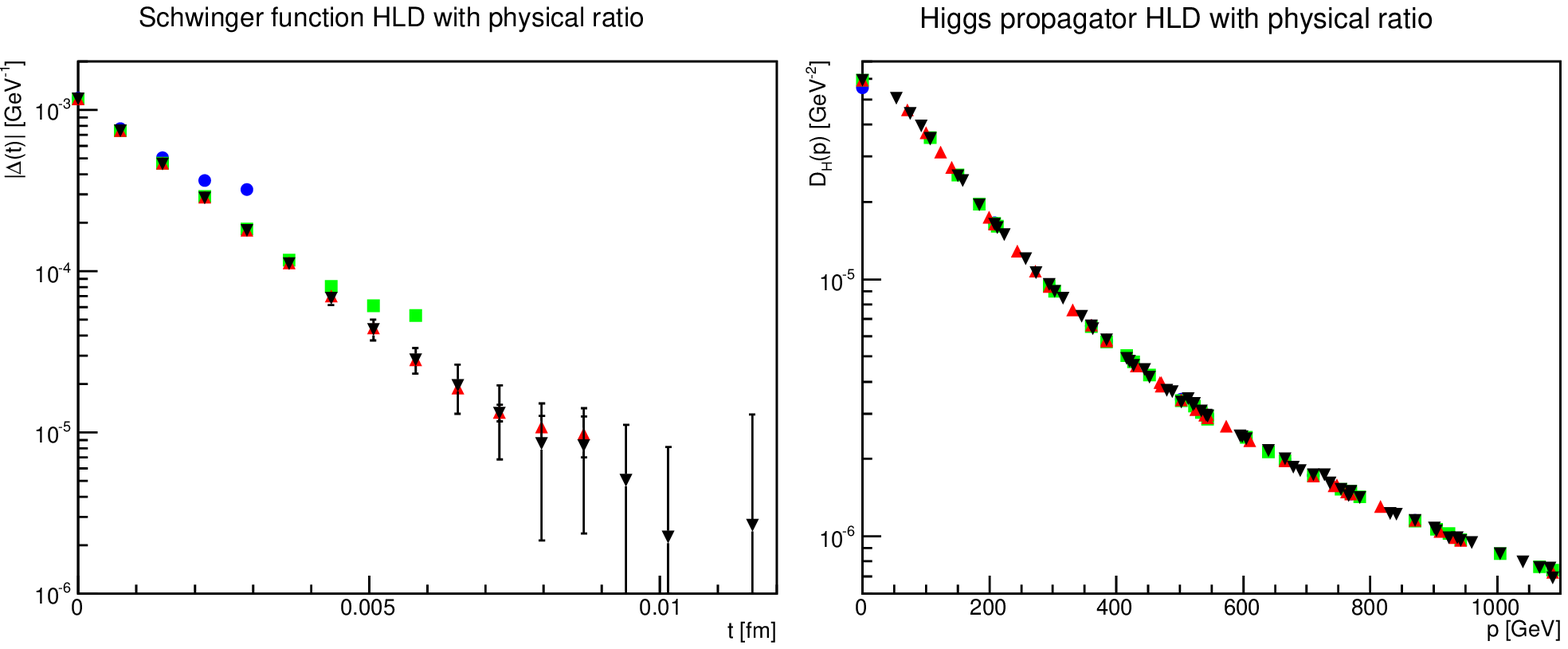}
\includegraphics[width=\linewidth]{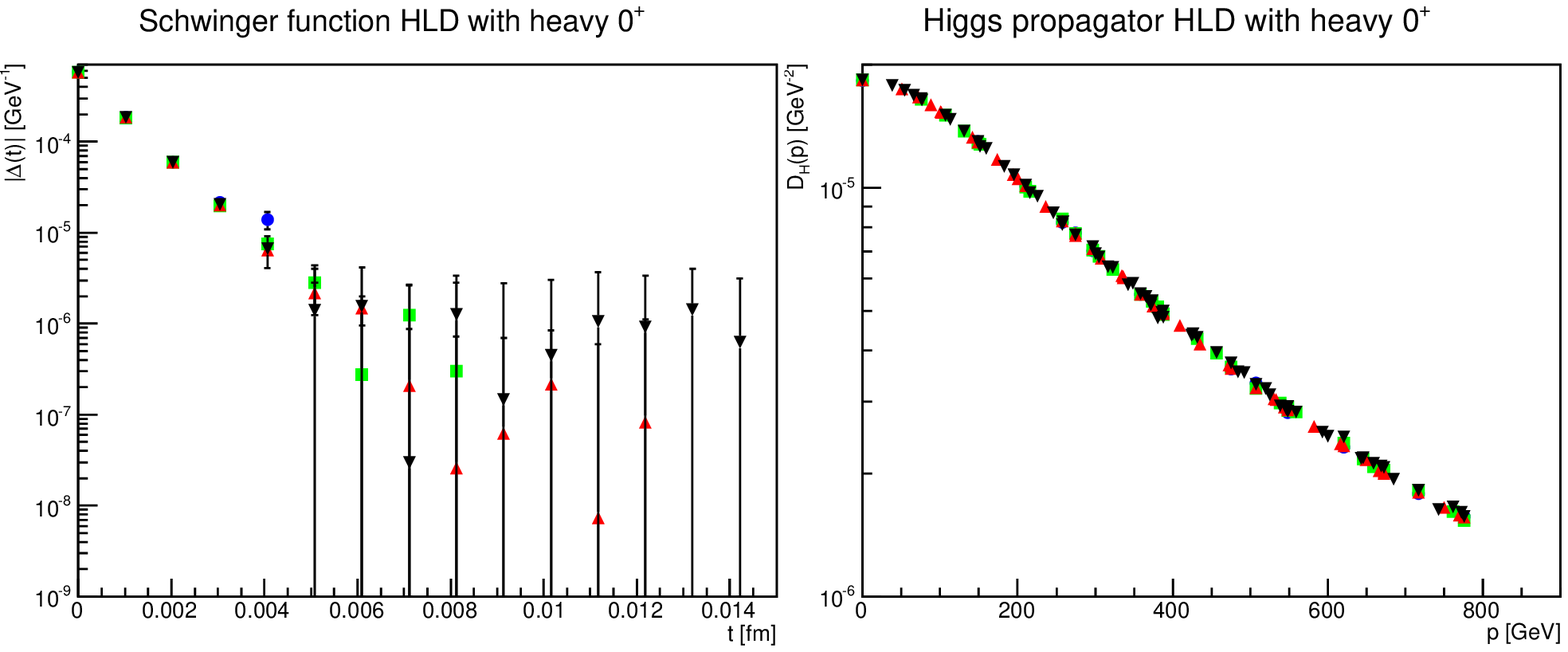}
\caption{\label{fig:hv}The Higgs propagator (right panel) and Schwinger function (left panel) for different volumes. The top panel is in the QLD with $m_{1^-}/m_{0^+}=2.2$, the middle panel has the physical mass ratio $m_{1^-}/m_{0^+}=0.72$, and the bottom panel is for a large Higgs mass $m_{1^-}/m_{0^+}=0.31$, both in the HLD. Note that the renormalization has been performed for all volumes with the same renormalization constants.}
\end{figure}

\begin{figure}
\centering
\includegraphics[width=\linewidth]{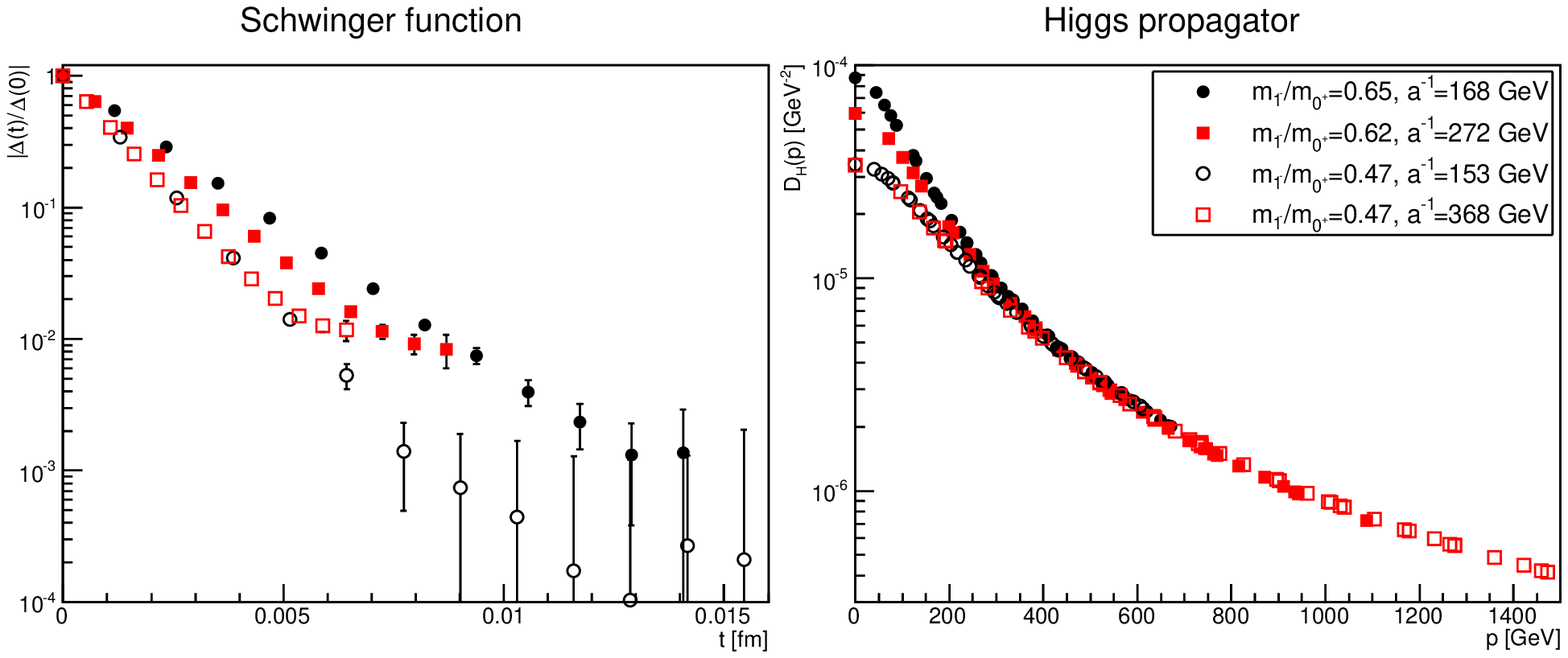}
\caption{\label{fig:ha}The Higgs propagator (right panel) and Schwinger function (left panel) for different mass ratios $m_{1^-}/m_{0^+}$ on 24$^4$ lattices and for two different lattice spacings.}
\end{figure}

The necessary repetition of the study of lattice artifacts for volume effects and discretization effects are shown in figures \ref{fig:hv} and \ref{fig:ha}, respectively. The first observation is that, in agreement with \cite{Maas:2010nc}, there is essentially no volume-dependence for the Higgs propagator, especially in comparison to the $W$ propagator. The same is true for the lattice spacing-dependency if the masses used for the renormalization purposes are truly identical. Otherwise the differing mass creates some difference. Nonetheless, in total the Higgs propagator is least affected by lattice artifacts.

\begin{figure}
\centering
\includegraphics[width=\linewidth]{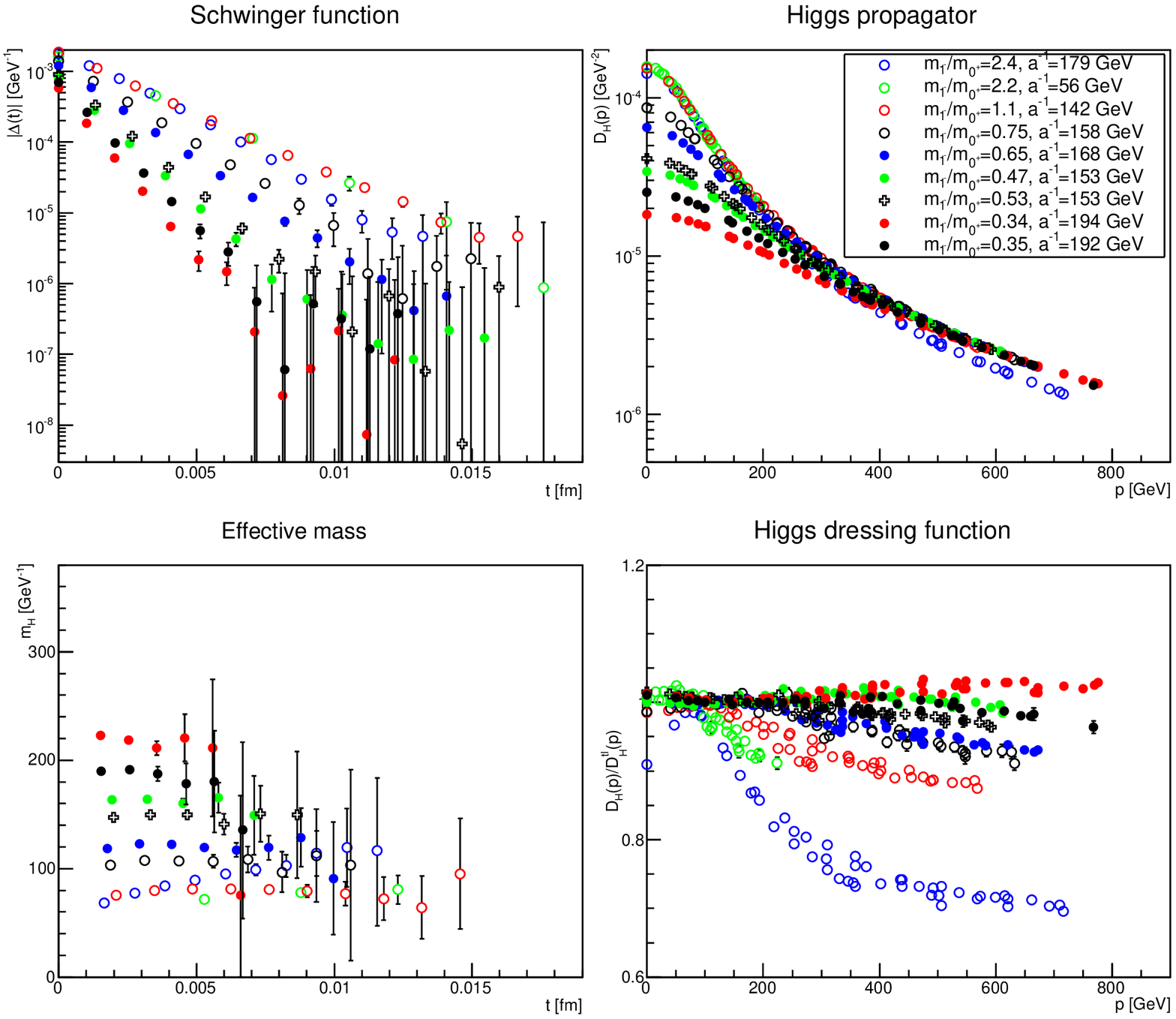}
\caption{\label{fig:h}The Higgs propagator (top-right panel), dressing function (bottom-right panel), Schwinger function (top-left panel), and effective mass (bottom-left panel) for different mass ratios $m_{1^-}/m_{0^+}$ on 24$^4$ lattices. All propagators are renormalized in the pole scheme, and $D_H^\tl=1/(p^2+m_H^2)$.}
\end{figure}

The results for the Higgs propagator for different mass ratios $m_{1^-}/m_{0^+}$ are shown in figure \ref{fig:h}. There are a number of intriguing observations. The first is that the propagators do not fully coincide at large momenta, even though being renormalized. This indicates that at least the effects of the mass, as a hard mass scale, pertain to larger momenta.

More intriguing is the behavior of the effective mass, which can already be inferred from the space-time correlator. In the QLD the effective mass bends upwards, signaling an unphysical behavior. This is not expected in this case in the same way as for the $W$ boson, since in the QLD the Higgs-like mass generation is not operative. Nonetheless, the Higgs shows also in the QLD at long times a behavior compatible with the mass induced by the renormalization prescription. In the HLD, however, the space-time correlator gets more and more into perfect agreement with an ordinary massive particle of the renormalized mass with increasing renormalized mass.

\begin{figure}
\centering
\includegraphics[width=\linewidth]{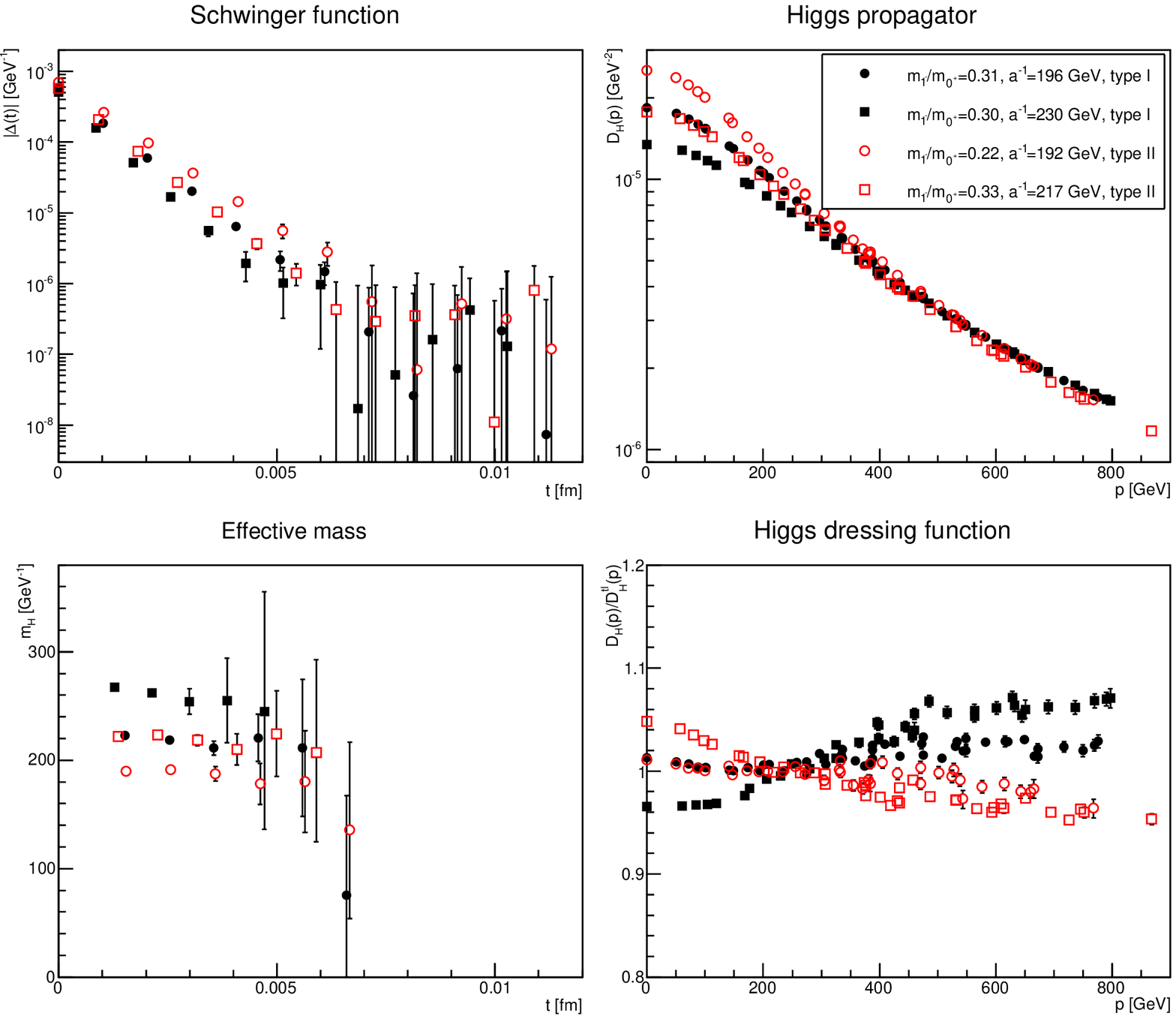}
\caption{\label{fig:hh}The Higgs propagator (top-right panel), dressing function (bottom-right panel), Schwinger function (top-left panel), and effective mass (bottom-left panel) for different mass ratios $m_{1^-}/m_{0^+}$ on 24$^4$ lattices. All propagators are renormalized in the pole scheme. Type I refers to situations where relation \pref{correl2} does not hold, while type II refers to situations where it does hold.}
\end{figure}

Only at large masses a surprising behavior sets in. At small Higgs masses, the propagator is decreasing faster than the tree-level one to which it is tied by the renormalization scheme \prefr{scheme1}{scheme2}, signaling the presence of the expected logarithmic corrections. This is the same behavior as in the quenched case \cite{Maas:unpublished,Maas:2011yx}. However, at small $m_{1^-}/m_{0^+}$ ratios, there appears a second behavior, in which it increases instead of decreasing. It appears that this is a systematic effect, which is tied to the validity of the relation \pref{correl2} for the $W$ boson, as can be seen in figure \ref{fig:hh}: The propagator decreases slower than tree-level if the relation \pref{correl2} is valid, and faster if the relation is violated. This behavior can actually be modified by choosing a different renormalization scheme, but the important observation here is that in a fixed scheme there is also for the Higgs propagator a possible difference between both cases.

Thus, at small $m_{1^-}/m_{0^+}$ ratios two different branches seem to appear, with distinct behaviors for the $W$ and the Higgs. This is not an effect of the running gauge coupling, where this behavior does not surfaces, but seems to be connected to the Higgs self-interaction. Concerning the corresponding bare parameters, the relation \pref{correl2} seems to be violated for a weaker Higgs self-interaction, in terms of the bare lattice parameters. This is also counter-intuitive. However, the number of such lattice parameter sets found is small so far. It appears necessary to significantly enlarge the sample, also over a wider range of $0^+$ masses and lattice spacings, before a conclusive statement can be made. It is, however, tempting to speculate that these two directions could manifest different kinds of physics when moving the lattice spacing to the minimum value possible. It is certainly a worthwhile endeavor to investigate this in more detail, also with respect to gauge-invariant physics \cite{Maas:unpublished3}.

\section{Vertices}\label{svertex}

\subsection{Ghost-$W$ vertex}

\begin{figure}
\centering
\includegraphics[width=\linewidth]{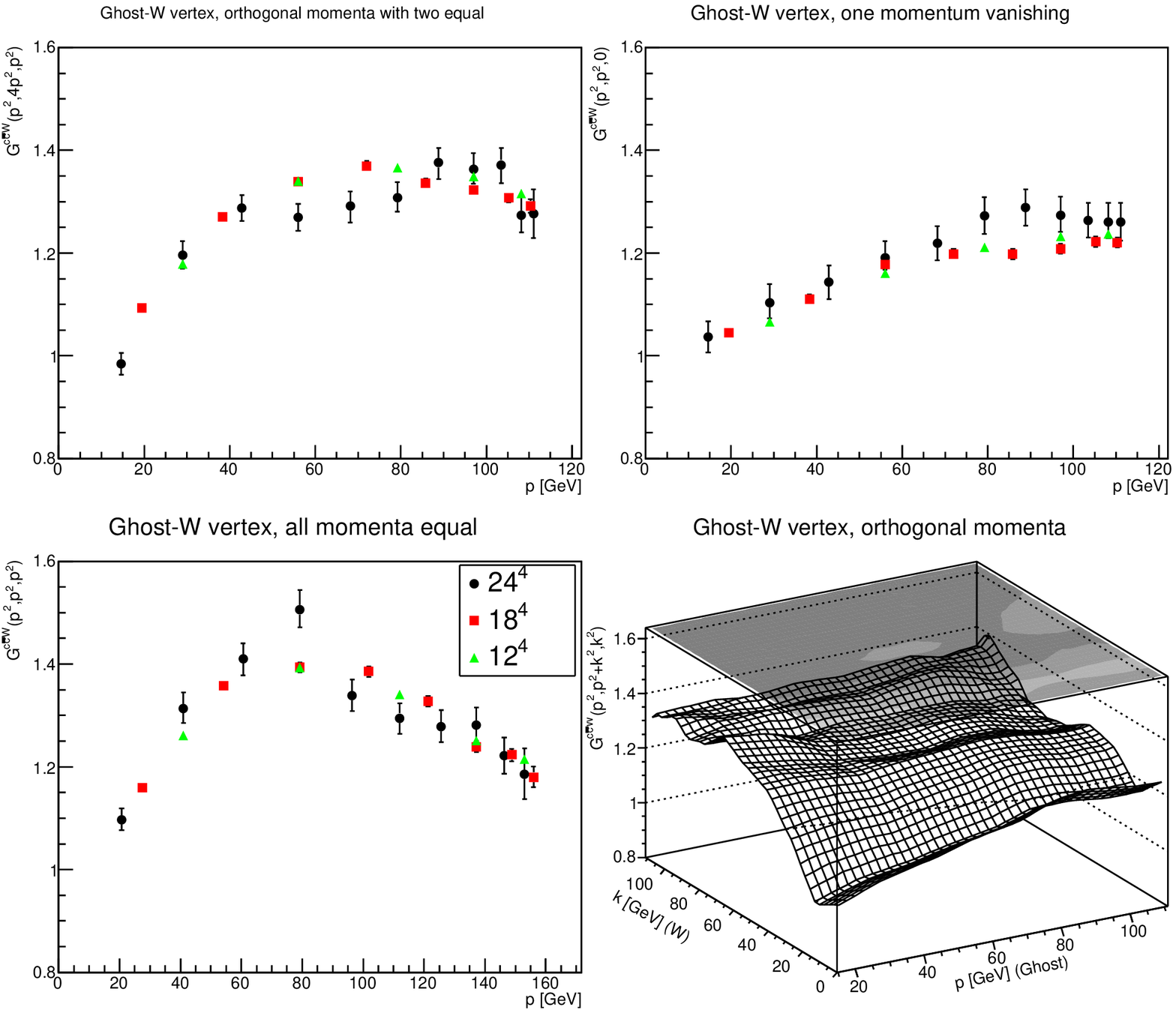}
\caption{\label{fig:ggv1}The ghost-$W$ vertex. The top-left panel shows the case of equal ghost and $W$ momentum, orthogonal to each other. The top-right panel shows the case for vanishing $W$ momentum. The bottom-left panel shows the symmetric configuration. The bottom-right panel is a three-dimensional plot of the possible ghost and $W$ momenta orthogonal to each other for the largest lattice volume. The mass ratio is $m_{1^-}/m_{0^+}=2.2$. The results are not renormalized.}
\end{figure}

\begin{figure}
\centering
\includegraphics[width=\linewidth]{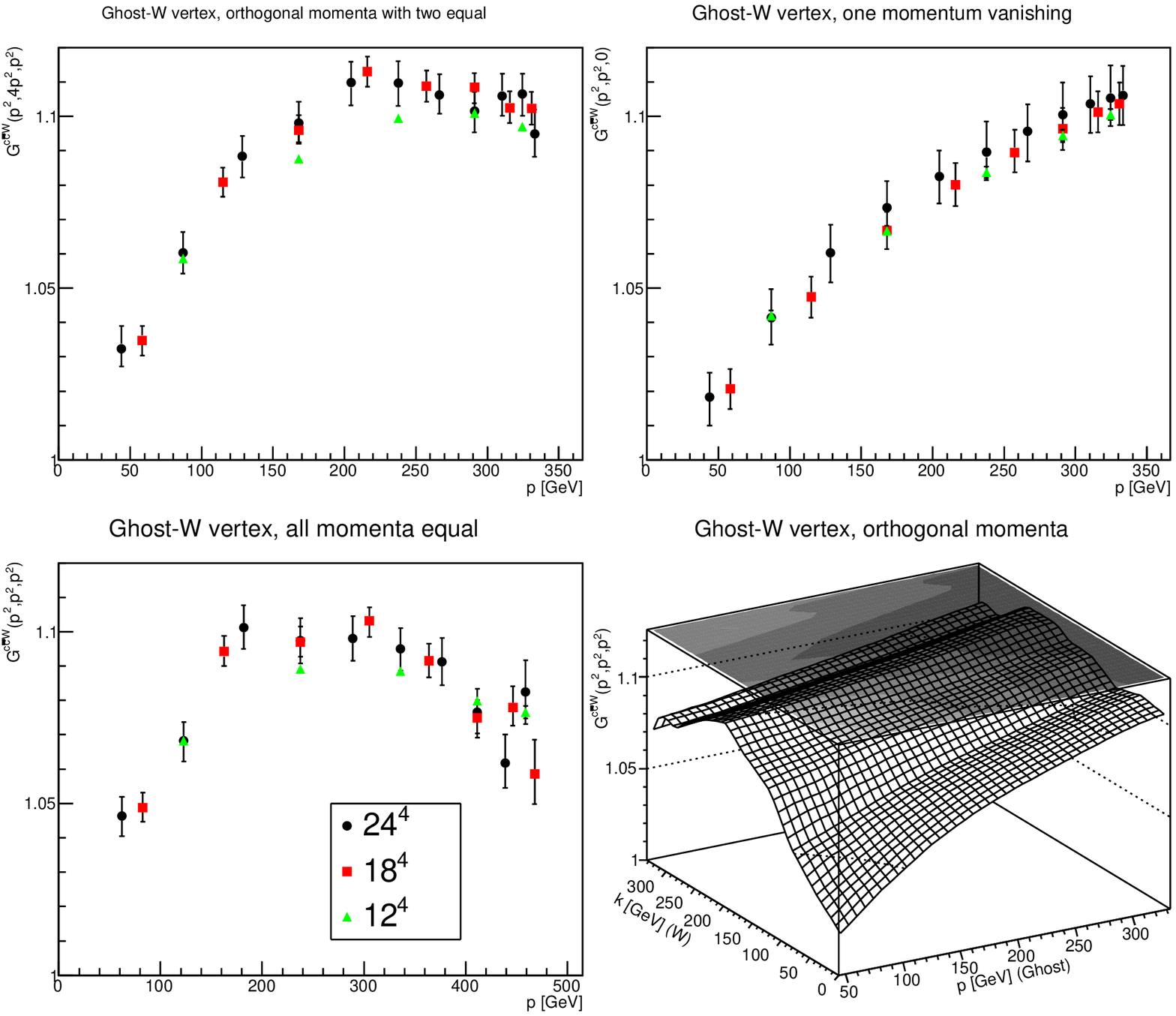}
\caption{\label{fig:ggv2}The ghost-$W$ vertex. The top-left panel shows the case of equal ghost and $W$ momentum, orthogonal to each other. The top-right panel shows the case for vanishing $W$ momentum. The bottom-left panel shows the symmetric configuration. The bottom-right panel is a three-dimensional plot of the possible ghost and $W$ momenta orthogonal to each other for the largest lattice volume. The mass ratio is $m_{1^-}/m_{0^+}=0.65$. The results are not renormalized.}
\end{figure}

\begin{figure}
\centering
\includegraphics[width=\linewidth]{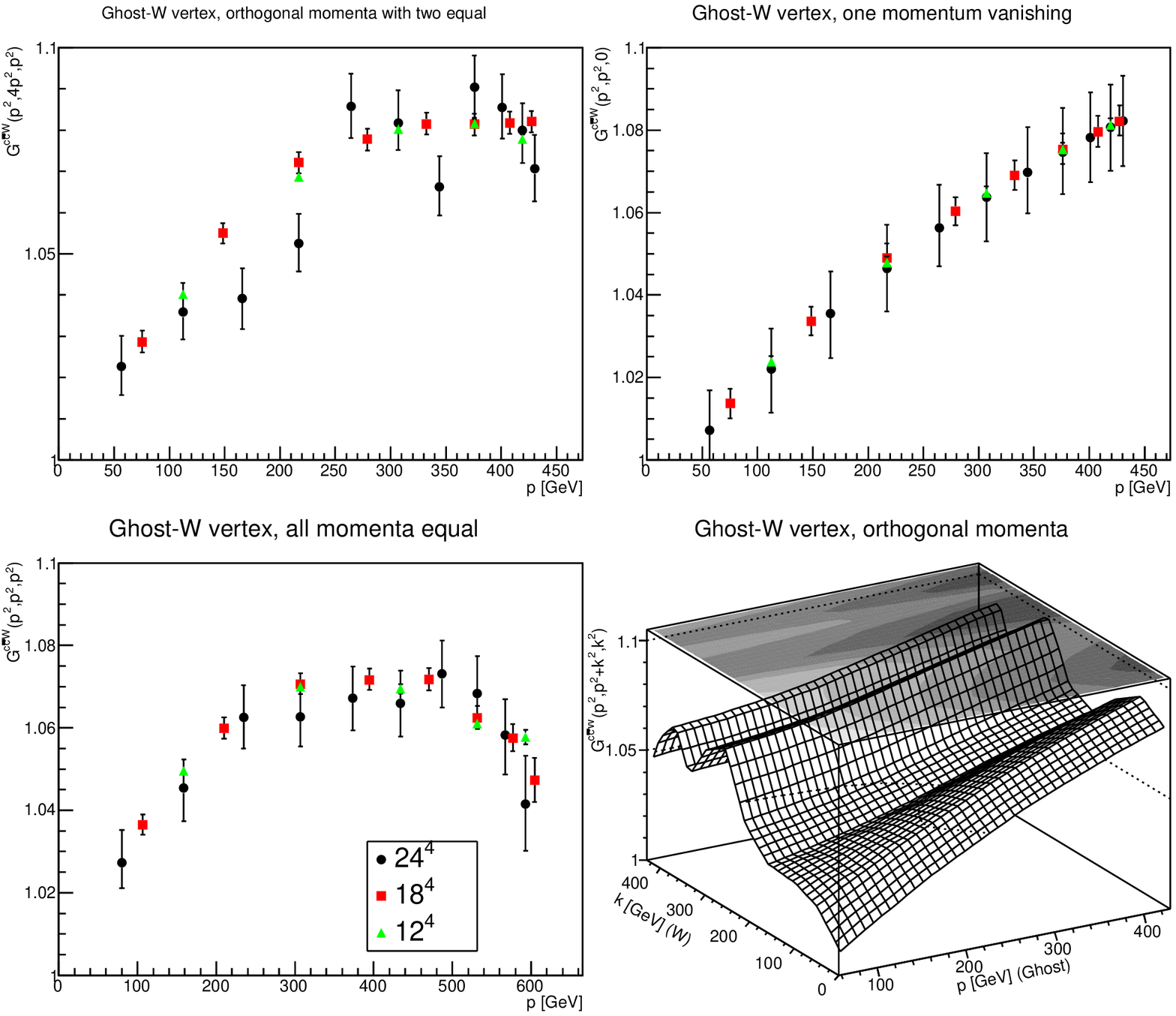}
\caption{\label{fig:ggv3}The ghost-$W$ vertex. The top-left panel shows the case of equal ghost and $W$ momentum, orthogonal to each other. The top-right panel shows the case for vanishing $W$ momentum. The bottom-left panel shows the symmetric configuration. The bottom-right panel is a three-dimensional plot of the possible ghost and $W$ momenta orthogonal to each other for the largest lattice volume. The mass ratio is $m_{1^-}/m_{0^+}=0.31$. The results are not renormalized.}
\end{figure}

The interaction three-point vertices are after the propagators the most simple objects, and the first objects which give insights into the interaction of the particles. The simplest, and statistically most simple one \cite{Cucchieri:2006tf}, is the ghost-$W$ vertex. It is shown for different mass ratios $m_{1^-}/m_{0^+}$ in figures \ref{fig:ggv1}-\ref{fig:ggv3}. It should be noted that in Landau gauge there is a ghost-anti-ghost symmetry \cite{Alkofer:2000wg}, and therefore the momentum-dependency for the anti-ghost momentum can be inferred from the one of the ghost.

Not surprisingly, given the results for the propagators, the vertex in the QLD, shown in figure \ref{fig:ggv1}, exhibits essentially the same behavior as in SU(2) Yang-Mills theory \cite{Maas:2011se,Cucchieri:2008qm,Huber:2012kd,Fister:2011ym,Pelaez:2013cpa}. Especially, the vertex is rather flat, except for a bump at an intermediate momentum of typical scale of the theory, here the mass of the lightest bound state.

The situation in the HLD for both a light $0^+$, shown in figure \ref{fig:ggv2}, as well as for a $0^+$ above threshold, shown in figure \ref{fig:ggv3}, is similar. The only difference is that the mid-momentum bump is severely reduced, and also shifted to larger masses of about two times the $0^+$ mass. Furthermore, the bump decreases with increasing $0^+$ mass. This could have also been inferred from the decrease and shift of the running coupling \pref{alpha} in figure \ref{fig:alpha}, as the relation \pref{alpha} stems from the relation between the ghost-$W$-vertex renormalization and the $W$ and ghost propagators \cite{vonSmekal:1997vx}.

\subsection{3-$W$-boson vertex}

\begin{figure}
\centering
\includegraphics[width=\linewidth]{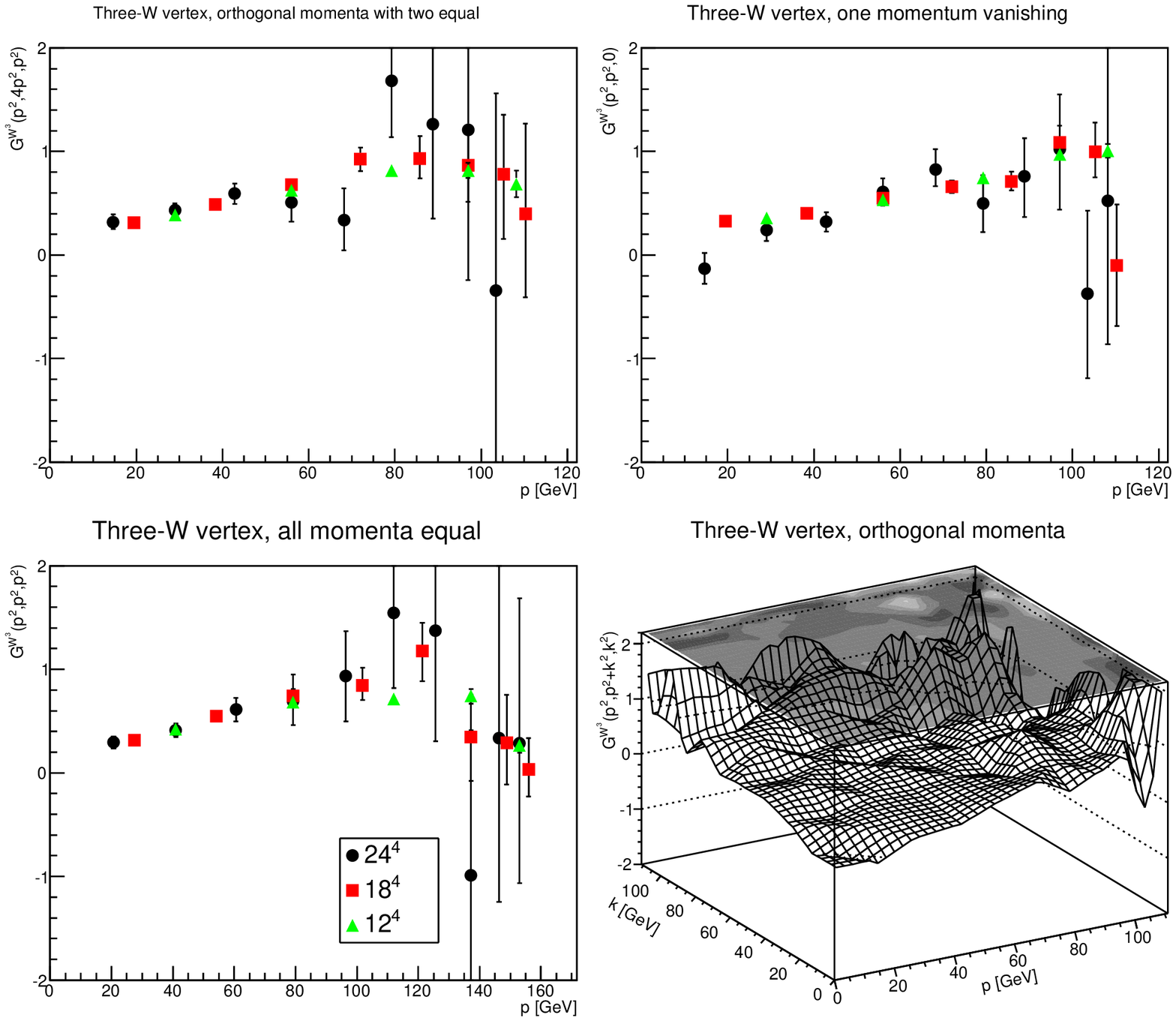}
\caption{\label{fig:g3v1}The three-$W$ vertex. The top-left panel shows the case of two equal $W$ momenta, orthogonal to each other. The top-right panel shows the case for one vanishing $W$ momentum. The bottom-left panel shows the symmetric configuration. The bottom-right panel is a three-dimensional plot of the possible ghost and $W$ momenta orthogonal to each other for the largest lattice volume. The mass ratio is $m_{1^-}/m_{0^+}=2.2$. The results are not renormalized.}
\end{figure}

\begin{figure}
\centering
\includegraphics[width=\linewidth]{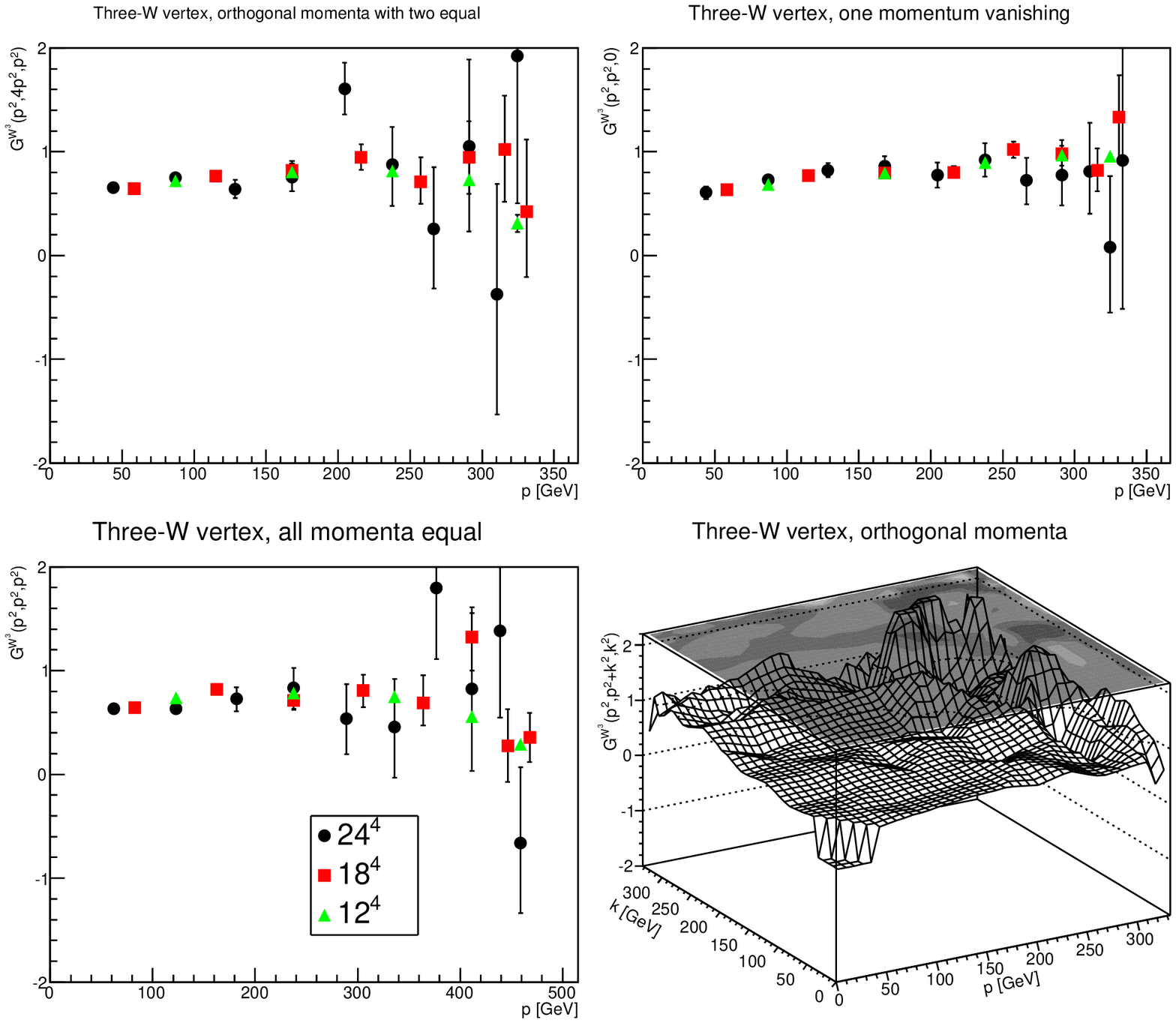}
\caption{\label{fig:g3v2}The three-$W$ vertex. The top-left panel shows the case of two equal $W$ momenta, orthogonal to each other. The top-right panel shows the case for one vanishing $W$ momentum. The bottom-left panel shows the symmetric configuration. The bottom-right panel is a three-dimensional plot of the possible ghost and $W$ momenta orthogonal to each other for the largest lattice volume. The mass ratio is $m_{1^-}/m_{0^+}=0.65$. The results are not renormalized.}
\end{figure}

\begin{figure}
\centering
\includegraphics[width=\linewidth]{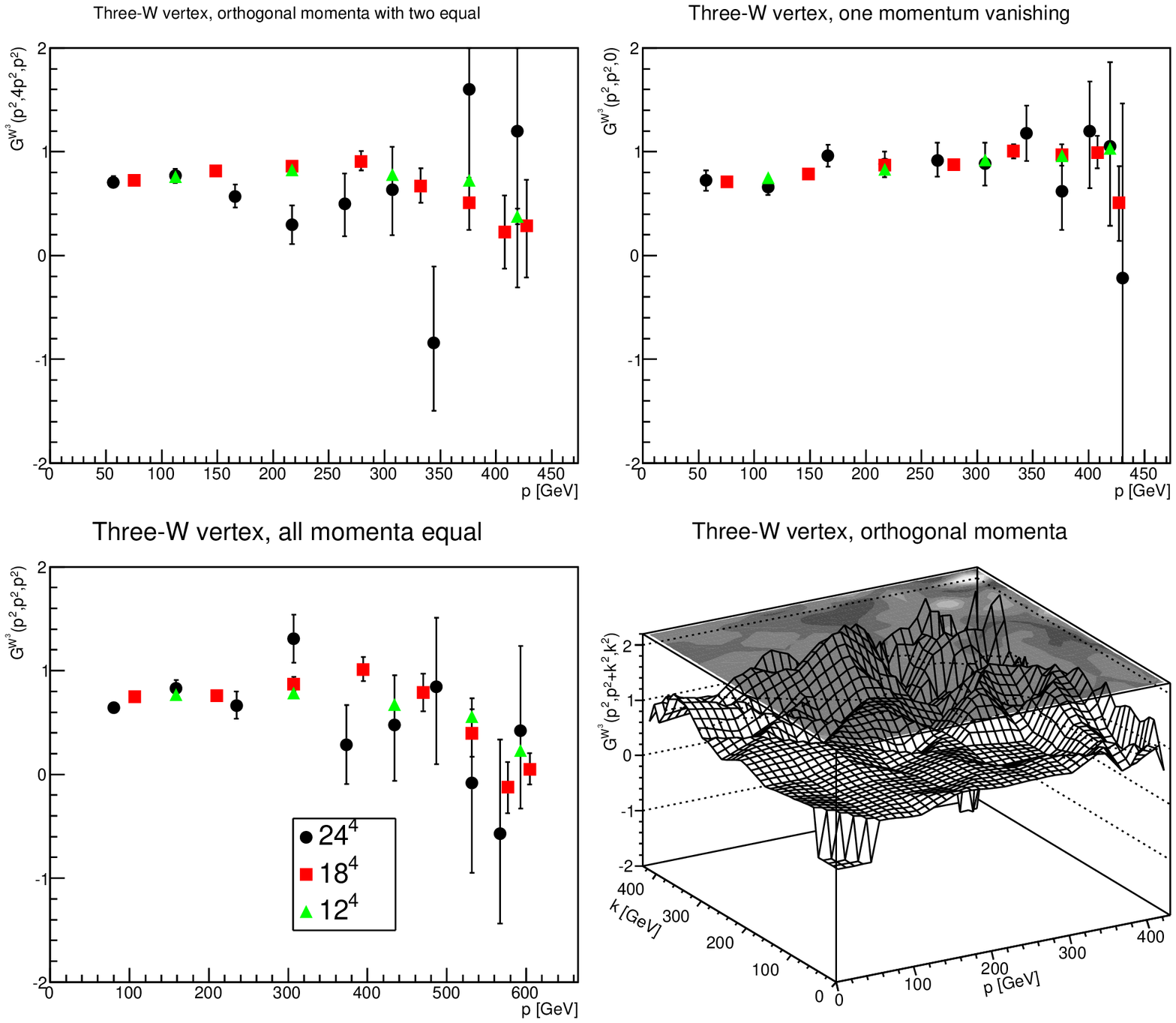}
\caption{\label{fig:g3v3}The three-$W$ vertex. The top-left panel shows the case of two equal $W$ momenta, orthogonal to each other. The top-right panel shows the case for one vanishing $W$ momentum. The bottom-left panel shows the symmetric configuration. The bottom-right panel is a three-dimensional plot of the possible ghost and $W$ momenta orthogonal to each other for the largest lattice volume. The mass ratio is $m_{1^-}/m_{0^+}=0.31$. The results are not renormalized.}
\end{figure}

The results for the 3-$W$ vertex, which is highly constrained due to the Bose symmetry of all legs, are shown in figures \ref{fig:g3v1}-\ref{fig:g3v3}. The results show, as in the Yang-Mills case \cite{Cucchieri:2008qm}, much stronger statistical fluctuations than for the ghost-$W$ vertex, especially at high momenta. This limits the reliability, especially for larger lattice volumes. At small momenta, however, the statistical noise is significantly smaller.

The QLD case is presented in figure \ref{fig:g3v1}. It shows the characteristic infrared suppression also seen in Yang-Mills theory \cite{Maas:2011se,Cucchieri:2008qm,Huber:2012kd,Pelaez:2013cpa}, and also compatible with a zero crossing at small momenta. However, just like in the Yang-Mills case in four dimensions \cite{Cucchieri:2008qm}, the volumes are just not large enough to unambiguously establish it. In the Yang-Mills case, the results in lower dimensions \cite{Cucchieri:2008qm,Maas:2007uv,Huber:2012td} clearly show this zero crossing, and it is therefore suggestive that this also should occur in four dimensions. The situation for the QLD here is very reminiscent of this. However, only larger volumes will finally permit to decide this question unequivocally.

The situation in the HLD, both for the low-mass $0^+$ in figure \ref{fig:g3v2} and the above-threshold $0^+$ mass in figure \ref{fig:g3v3}, is somewhat different. Here, the results do not show a strong tendency for an infrared suppression, though a slight decrease is observed. Still, the results extrapolate much better to a finite value. However, in units of the lightest excitation, the volumes in both cases are substantially smaller than for the QLD calculation. This may therefore be a finite volume effect.

Much clearer is that there is little, if at all, dependency on the mass of the $0^+$, at least within the errors. It will require more systematic investigations at larger volumes to clarify the behavior in the HLD.

\subsection{Higgs-$W$ vertex}

The last vertex is, in principle, the most interesting one, the $W$-Higgs vertex. Not only because it is the mediator of the Higgs effect \cite{Bohm:2001yx}, but it is also suspected to play an important role in the confinement process in the QLD \cite{Fister:2010yw}. Unfortunately, and somewhat surprisingly, it is even stronger affected by statistical fluctuations than the 3-$W$ vertex. This made a large-volume study of it at the current time essentially not feasible. Here, the results, as far a possible are presented, though the large statistical uncertainty beyond the smallest volume make the results only of limited systematic reliability.

\begin{figure}
\centering
\includegraphics[width=\linewidth]{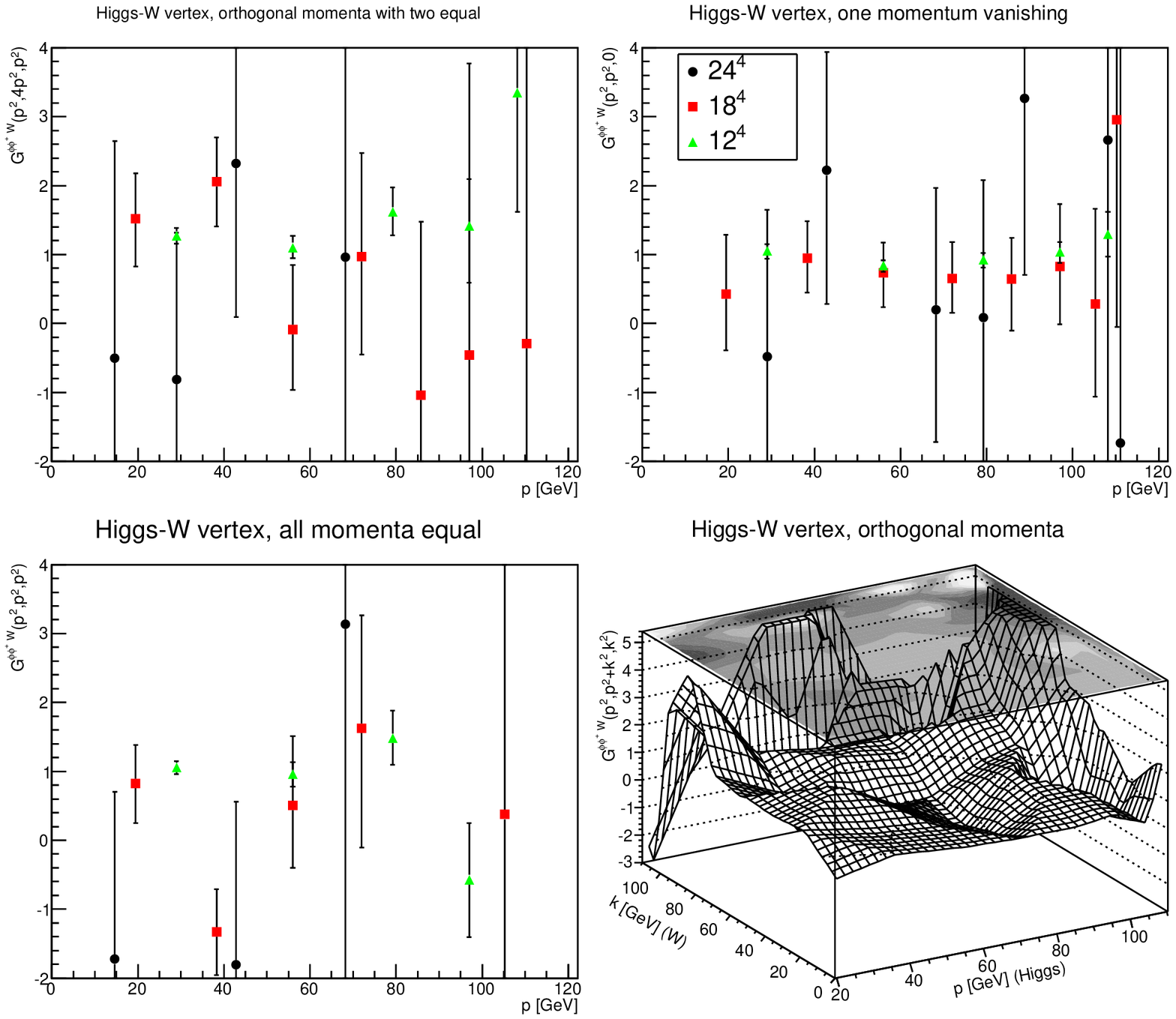}
\caption{\label{fig:ghv1}The Higgs-$W$ vertex. The top-left panel shows the case of equal Higgs and $W$ momentum, orthogonal to each other. The top-right panel shows the case for vanishing $W$ momentum. The bottom-left panel shows the symmetric configuration. The bottom-right panel is a three-dimensional plot of the possible ghost and $W$ momenta orthogonal to each other for the $18^4$ lattice. The mass ratio is $m_{1^-}/m_{0^+}=2.2$. The results are not renormalized.}
\end{figure}

\begin{figure}
\centering
\includegraphics[width=\linewidth]{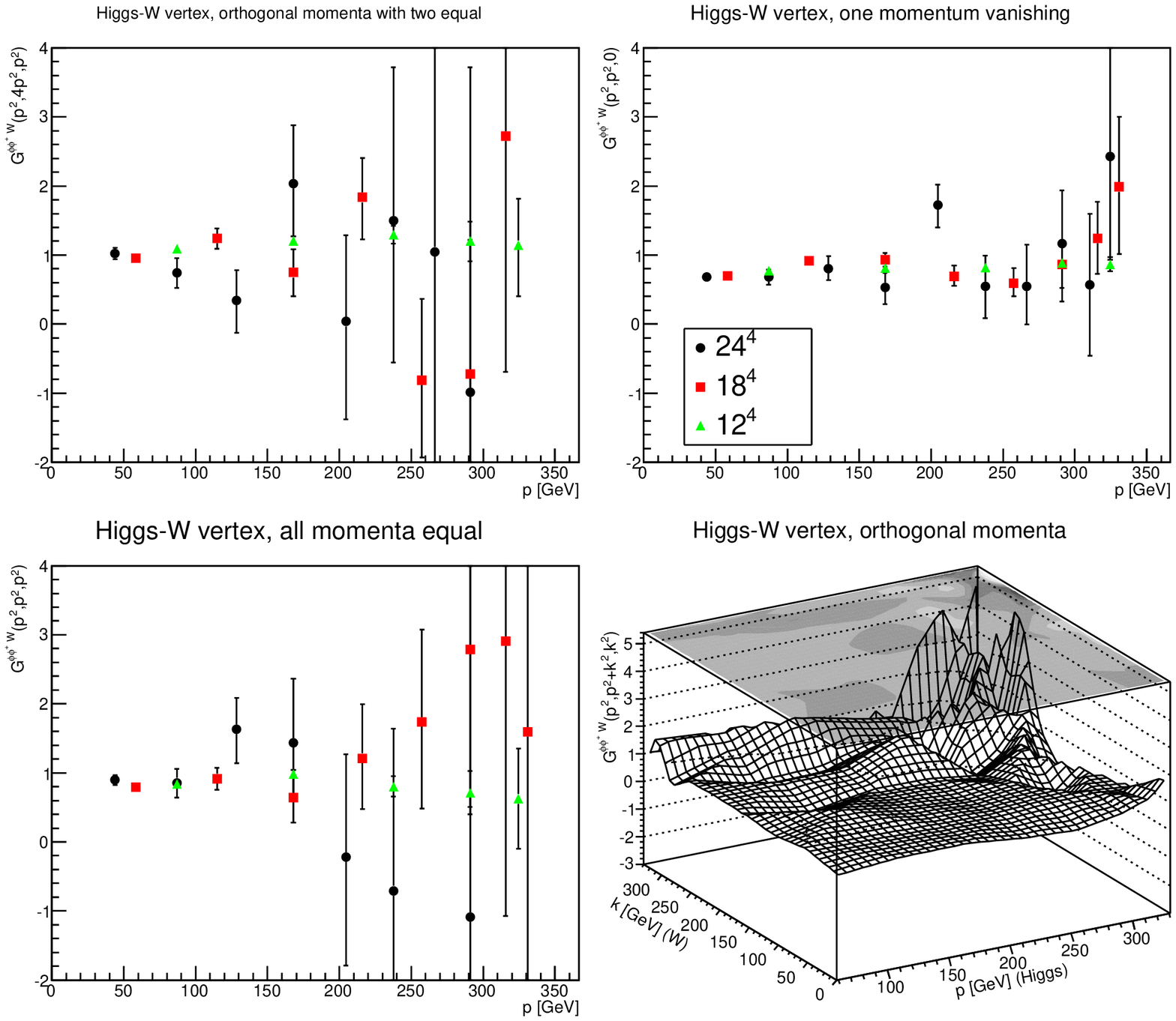}
\caption{\label{fig:ghv2}The Higgs-$W$ vertex. The top-left panel shows the case of equal Higgs and $W$ momentum, orthogonal to each other. The top-right panel shows the case for vanishing $W$ momentum. The bottom-left panel shows the symmetric configuration. The bottom-right panel is a three-dimensional plot of the possible ghost and $W$ momenta orthogonal to each other for the $18^4$ lattice. The mass ratio is $m_{1^-}/m_{0^+}=0.65$. The results are not renormalized.}
\end{figure}

\begin{figure}
\centering
\includegraphics[width=\linewidth]{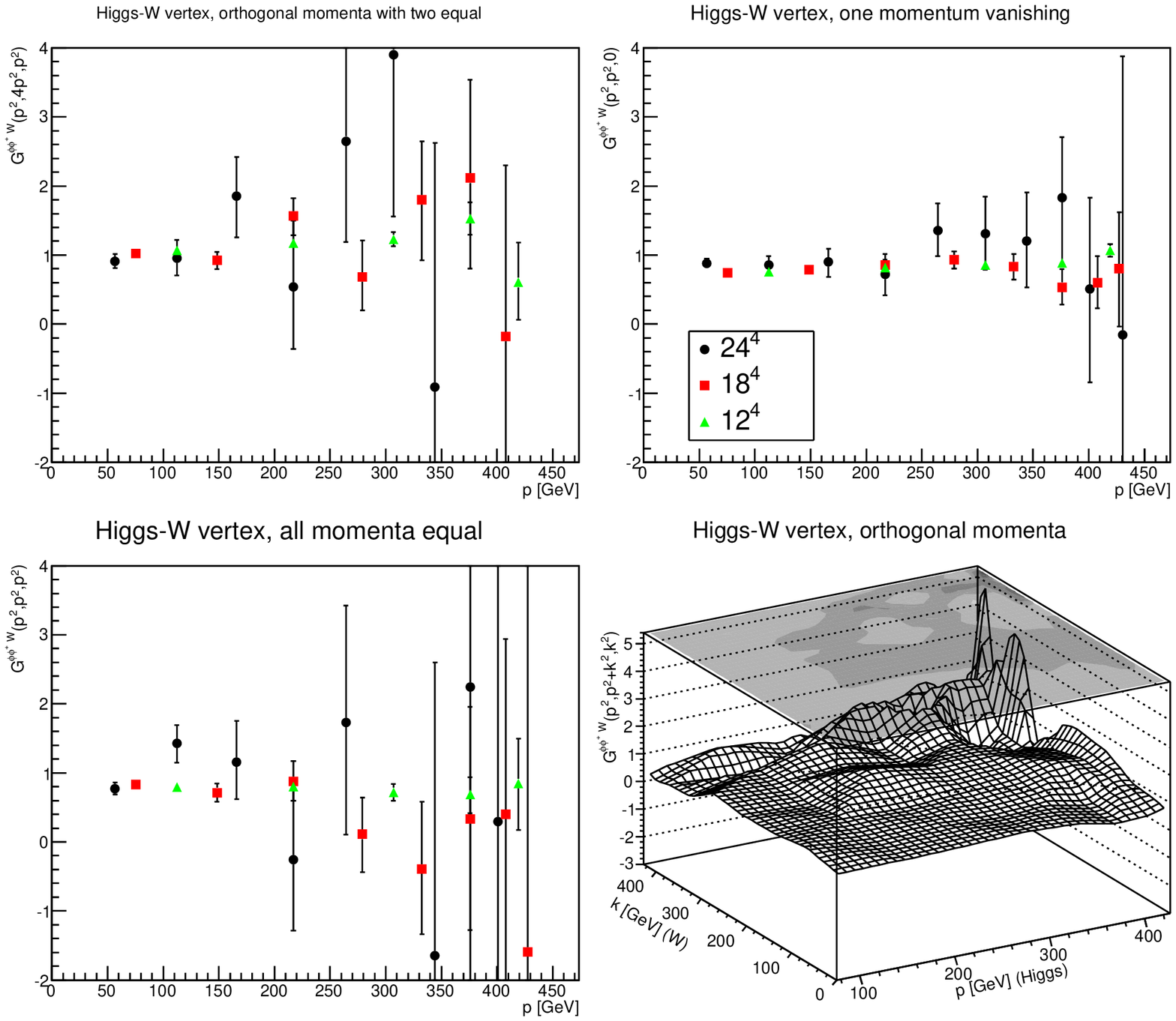}
\caption{\label{fig:ghv3}The Higgs-$W$ vertex. The top-left panel shows the case of equal Higgs and $W$ momentum, orthogonal to each other. The top-right panel shows the case for vanishing $W$ momentum. The bottom-left panel shows the symmetric configuration. The bottom-right panel is a three-dimensional plot of the possible ghost and $W$ momenta orthogonal to each other for the $18^4$ lattice. The mass ratio is $m_{1^-}/m_{0^+}=0.31$. The results are not renormalized.}
\end{figure}

The dressing functions are shown in figure \ref{fig:ghv1}-\ref{fig:ghv3}. The statistical fluctuations are worst in the QLD, shown in figure \ref{fig:ghv1}, and decrease with increasing $0^+$ mass in the HLD, i.\ e.\ from figure \ref{fig:ghv2} to figure \ref{fig:ghv3}. The results are compatible with a more or less flat momentum behavior, though the differences between the case with vanishing $W$ momentum and non-vanishing $W$ momentum for the orthogonal configurations are compatible with some angular dependence on the angle between the Higgs and the anti-Higgs. Since in the present case there is no symmetry between the two legs, this is not excluded.

Such an essentially flat behavior is also compatible with the quenched case, though there no significant angular dependence is observed \cite{Maas:2011yx,Maas:unpublished}. The results are furthermore not compatible with any kind of divergence, either towards the infrared, nor towards vanishing $W$ momentum, i.\ e.\ of any kind of kinematical singularity. This is the case in both the QLD and the HLD, and appears to preclude any possibility to obtain a strong contribution to the intermediate distance string tension from a single $W$ exchange, as has been discussed for QCD \cite{Alkofer:2008tt,Alkofer:2000wg}.

\subsection{A note on the four-point vertices}\label{s4p}

The previously shown results indicate that the Higgs can have quite an impact on the gauge boson, in stark contrast to the case of fermions with the same number of degrees of freedom, even when freely varying their mass. It appears therefore possible that the Higgs-self-interaction plays an important role in this context, since this is already the case at the classical level \cite{Bohm:2001yx}. Unfortunately, the Higgs-self-coupling makes its first direct appearance in this gauge at the level of the four-point functions.

In the present gauge there are six such four-point functions with the generic structure $\langle B_a \bar{B}_b B_c \bar{B}_d\rangle$, with collective indices including field type and $\bar{B}$ is the anti-particle, which is identical to the particle in case of the $W$ field. These are the ghost-ghost scattering kernel, the ghost-$W$ scattering kernel, the ghost-Higgs scattering kernel, the $W$-$W$ scattering kernel, the $W$-Higgs scattering kernel, and the Higgs-Higgs scattering kernel. There are two main issues with the calculation of these four-point functions.

One is that the amount of statistical fluctuations will be larger than the one for the corresponding three-point functions, especially the larger the number of Higgs fields, given the comparison between the three-$W$ and the $W$-Higgs vertex above. The second is that in the non-aligned Landau gauge these are the first correlation functions for which connected and full correlation functions do not agree, but disconnected contributions have to be removed,
\be
\langle B_a \bar{B}_b B_c \bar{B}_d\rangle_\text{connected}=\langle B_a \bar{B}_b B_c \bar{B}_d\rangle-\sum_P c_P \langle B_{a_P}\bar{B}_{b_P}\rangle \langle B_{c_P}\bar{B}_{d_P}\rangle 
\ee
\no where the sum is over permutations of the indices and $c_P$ is a constant depending on the involved field types. This increases the required statistical precision even further, pushing these objects out of our numerical reach, as noted in the introduction. The only possible exception may be the ghost-ghost scattering kernel, since due to the inversion of the Faddeev-Popov operator and the therefore included lattice averaging it is less affected by statistical fluctuations.

There is one further exception. For the case of the Higgs-Higgs scattering kernel, there is a gauge-invariant contraction of the indices, if the arguments of the Higgs and the anti-Higgs fields pairwise coincide. This is then just the Higgsonium operator \pref{higgsonium}. Since no gauge-fixing is required to determine it, this channel can be statistical accessed with sufficient brute force \cite{Maas:unpublished3,Jersak:1985nf,Langguth:1985eu}, and at least its pole structure can be accessed, giving the physical excitations in the $0^+$ channel. The relation \pref{correl} shows also that, for a physical Higgs mass, there is a connection to the perturbative one-Higgs exchange in this channel in an aligned gauge, which is, e.\ g.\ absent in the QLD, where the dominant part will be a two-Higgs exchange. Thus, the relation \pref{correl} already implies that the Higgs-Higgs scattering kernel will exhibit at least one perturbative feature. This makes it even more interesting to understand which role it plays in the influence of the Higgs on the gauge sector. However, this will have to await significant more computational resources, or different approaches, like, e.\ g., functional methods \cite{Maas:2011se}.

Note that no such argument can be made in case of the $W$-$W$ scattering kernel, as the simplest gauge-invariant objects formed only from $W$ fields involves at least eight $W$ fields, the plaquette and the topological charge density.

\section{Conclusions}\label{sconclusions}

Summarizing, we have presented an extensive study of two-point functions and, for the first time, three-point functions in Yang-Mills-Higgs theory in the non-aligned minimal Landau gauge using lattice methods throughout a significant part of the phase diagram of the theory.

We have confirmed earlier results \cite{Langguth:1985eu,Evertz:1985fc} that the theory undergoes a drastic change from a would-be Higgs behavior to a would-be QCD behavior when the mass of the $0^+$ drops below the one of the $1^-$ state from the investigation of these correlation functions. Of course, this is true only away from the overlap region, where the transition is a cross-over and many aspects become gauge-dependent \cite{Caudy:2007sf}. But already quite close by this cross-over the correlation functions show a pronounced QCD-like or Higgs-like behavior, especially visible in the gauge sector. Inside this QCD-like region the correlation functions in the gauge sector show a behavior close to the one of Yang-Mills theory \cite{Maas:2011se}, while the ones involving Higgs fields are close to the quenched case \cite{Maas:2011yx}. These results are in line with most expectations from functional studies \cite{Gies:2013pma,Fister:2010yw,Macher:2011ad,Fischer:2009tn,Hopfer:2013via,Mitter:2013me,Capri:2012ah}, and proposals which involve infrared divergent $W$-Higgs vertices \cite{Fister:2010yw} appear currently rather unlikely.

We have furthermore extended the observations from \cite{Maas:2012tj} and confirmed that the relations \prefr{correl}{correl2} established in \cite{Frohlich:1981yi} hold true as long as the $0^+$ is below the threshold for decays into two $1^-$. In this region, the propagators and vertices are close to the ones of perturbation theory \cite{Bohm:2001yx}. Especially, the Higgs and the $W$ are both massive, though the latter changes gradually into a massless particle at high energies. At these large energies they therefore coincide with the ones of the QCD-like domain.

If the mass of the $0^+$ exceeds twice the mass of the $1^-$, i.\ e.\ when it crosses the threshold for decays, the situation changes. Especially, two different behaviors are observed, which depend on the relative sizes of the bare lattice parameter. Note that this is not dependent on the running gauge coupling, which is found just to diminish continuously with increasing $0^+$ mass. The behavior observed is either a branch where the relations \prefr{correl}{correl2} do no longer hold, i.\ e.\ perturbation theory is no longer an adequate description. The other branch still shows this behavior, but the Higgs propagator shows at short distances no longer a behavior compatible with a simple massive particle. Hence, in both cases something interesting occurs. To fully understand the effect, this will require much more systematic investigations, as well as a comparison to the gauge-invariant physics of this part of the phase diagram, which will be done elsewhere \cite{Maas:unpublished3}.

Still, it seems to be likely that the simple perturbative behavior is at least valid in the region $1/2\le m_{1^-}/m_{0^+} \le 1$, in which the physical Higgs mass resides.\\

\no{\bf Acknowledgments}

This project was supported by the DFG under grant number MA 3935/5-1. T.\ M.\ was also supported by the DFG graduate school GRK 1523/1. Simulations were performed on the HPC cluster at the University of Jena. The authors are grateful to the HPC team for the very good performance of the cluster. The ROOT framework \cite{Brun:1997pa} has been used in this project.

\appendix

\section{Some remarks on variables and gauges}\label{a:remarks}

\subsection{Gauge-invariant variables}\label{a:giv}

The Yang-Mills-Higgs theory with two flavors of Higgs fields has a very interesting property \cite{Frohlich:1981yi}, which has been used repeatedly in lattice calculations \cite{Langguth:1985eu,Philipsen:1996af}: It is possible to rephrase the lattice action entirely in terms of the gauge-invariant operators describing the $0^+$ and $1^-$ excitations \pref{higgsonium} and \pref{wl}, the latter with $\alpha=-1$. By this the integration over the gauge orbit factorizes, and can be removed.

The price to be paid is twofold. One is that the topological structure of the target space changes from ${\cal R}^4$ to $SU(2)\times{\cal R}_+$, and is therefore partly compactified. Though such a change of target space does not seem to influence pertinent features in the ungauged case, like triviality \cite{Callaway:1988ya,Kenna:1993fp,Fernandez:1992jh}, it is not entirely clear whether this holds true for the gauged case, in which also the gauge fields offer non-trivial topological structure.

Aside from this more fundamental point, this change of variables entails a non-trivial Jacobian, which essentially manifests in form of an additional term $\ln\rho$ on the level of the Lagrangian \cite{Langguth:1985eu}
\bea
S&=&\beta\sum_{x\mu<\nu}\left(1-\frac{1}{2}\Re\tr V_{\mn}(x)\right)\nn\\
&&+\sum_x\left(\rho^2(x)-3\log\rho(x)+\lambda(\rho(x)-1)^2-\kappa\sum_{\mu>0}\rho(x+\mu)\rho(x)\tr V_\mu(x)\right)\nn
\eea
\no where $V_{\mn}$ is the plaquette obtained from the $V_\mu$. Note that this theory only retains the global flavor symmetry, as the last term would no longer be invariant under local gauge transformations. Thus, already at tree-level, an infinite number of vertices appear due to the $\ln\rho$ term, and perturbative renormalizability becomes quite difficult to achieve, if possible at all.

Of course, this poses no problem for lattice calculations, but so neither does a formulation including the gauge fields. If this additional term can be neglected perturbatively, this formulation has turned out to be quite useful \cite{Fodor:1994sj}.

If the fields are coupled, like in the standard model, to other gauge interactions, these variables are, of course, no longer gauge-invariant. Hence, their use is somewhat limited on a conceptual level, despite their technical usefulness. This approach is therefore not pursued here. Furthermore, there is some problem when the Higgs field vanishes, as then the action becomes locally infinite, as the Jacobian becomes singular.

\subsection{Unitary gauge}

One particular convenient way of gauge-fixing at tree-level in this theory is superficially unitary gauge \cite{Bohm:2001yx}, see e.\ g.\ \cite{Zubkov:2010np}. In this gauge, on each gauge orbit the gauge copy is chosen for which the $\varphi^a$ become unit matrices. Since a gauge transformation $g$ achieving this is given by $\varphi^{a-1}$, this is in general possible, since $\varphi$ is almost everywhere a valid SU(2) group element. However, at those remaining points, i.\ e.\ those at which the Higgs field $\phi$ vanishes, this gauge transformation is ill-defined, i.\ e.\ gauge defects are introduced. In contrast to the Landau gauge used in the main part of the text, it is therefore not a fully well-defined gauge, though this is of little importance on a finite lattice.

There are also further disadvantages. One is that again this changes the topology of the target space of the Higgs field. The second is that this gauge is perturbatively non-renormalizable at the level of gauge-dependent correlation functions \cite{Bohm:2001yx}, entailing problems in defining the correlation functions of the $W$ and the Higgs.

Formally, when writing down the corresponding gauge-fixed operators for the $W$ and Higgs field, these are in fact identical to the ones obtained when making the choice of gauge-invariant variables in the previous section \ref{a:giv}, i.\ e.\ \pref{higgsonium} and \pref{wl}. The main difference in practical terms is hence that in the previous case the transformation is done before evaluating the path integral, while in the latter case rather a $\delta$-functional
\be
\delta\left(\varphi^a-1\right)\nn,
\ee
\no as the gauge condition is introduced into the path integral. Thus, at the conceptual level, previously the points of vanishing Higgs field yield an infinite action, while they appear as gauge-fixing defects in the present case. Hence, aside from these points both approaches are equivalent. However, while the change of variables ceases to yield a gauge-invariant formulation when adding additional fields, and therefore is no longer useful, unitary gauge remains a gauge even in that case.

\subsection{'t Hooft gauge}

To avoid the problems introduced by the perturbative non-renormalizability of unitary gauge, perturbative calculations usually employ gauges like the 't Hooft gauges with the gauge condition \cite{Bohm:2001yx}
\be
\pdm A_\mu^a+i\zeta \phi_i\tau^a_{ij} n_j v=0\nn.
\ee
\no where $\zeta$ is a second gauge parameter, which is in general different from the gauge parameter $\xi$ of the covariant part of the gauge fixing. Usually, however, renormalization schemes are employed which ensure $\xi=\zeta$ to avoid mixing between Goldstone bosons and gauge fields \cite{Bohm:2001yx}. Only this version will be discussed here.

It is, of course, possible to take the limit $\xi=\zeta\to 0$, in which case the resulting gauge is the Landau gauge. However, for every non-vanishing value of the gauge parameters, the masses of the $W$ boson and the Higgs remain unchanged, while the masses of the ghosts and the Goldstones go with the gauge parameters to zero \cite{Bohm:2001yx}. In contrast, if instead of taking the limit, the gauge parameters are just set to zero, not only the Goldstones and ghosts will have vanishing mass, but so will the $W$ boson and the Higgs mass becomes tachyonic. These statements hold true to all orders in perturbation theory, except for the Higgs mass. Hence, while the limit is perturbatively well-defined, the situation at zero, which is the one employed in this work, is perturbatively not well-defined \cite{Lee:1974zg}. Non-perturbatively, these gauges are still well-defined. The gauge condition plays hence the role of an external magnetic field, which forces during the limiting process the system into a preferred vacuum, while the system at zero gauge parameter remains (classically) in the metastable symmetric situation \cite{Maas:2012ct}. From the point of view of non-perturbative calculations, however, this does not matter, and any choice is equally well possible.

Hence, as to be expected in a situation with metastability, taking the end-point of the sequence is not a continuous part of the sequence itself.

\section{Perturbation theory}\label{a:pt}

This still entails the question of how the results of the present work can be compared to perturbative calculations, and thus whether the statements about the validity of perturbation theory are reliable. Here helps the fact that the limit of 't Hooft gauge and Landau gauge only differ by averaging over the global part of the gauge group \cite{Maas:2012ct}. Hence, all quantities which are invariant under global gauge transformations remain invariant. Especially, this implies that though the individual components of the propagators are not invariant under a global gauge transformations, their traces are. Since here only such traces are calculated, these results will coincide in both gauges. Hence, they can be perturbatively calculated in the limit of 't Hooft gauge.

Furthermore, since the global gauge symmetry is explicitly manifest in the present gauge, all off-diagonal elements of propagators will vanish, and all diagonal elements are identical. This finally permits to determine the full propagators. Especially, this implies that at tree-level the propagators will behave as \cite{Bohm:2001yx}
\bea
D^{ab}&=&\delta^{ab}\left(\delta_\mn-\frac{p_\mu p_\nu}{p^2}\right)\frac{1}{p^2+m_W^2}\nn\\
D_G^{ab}&=&-\delta^{ab}\frac{1}{p^2}\nn\\
D^{ij}_H&=&\delta^{ij}\frac{1}{p^2+m_H^2}\nn,
\eea
\no where $m_W$ and $m_H$ are the corresponding tree-level masses.

At sufficiently large energy, the consequences of the Higgs effect quickly diminish, and therefore the propagators decay like massless particles, proportional to $\ln^{\delta_i} p^2$, with the relevant anomalous dimensions $\delta_i$, which can be obtained from resummed perturbation theory. However, because of the relation \pref{alpha} and the renormalization of the propagators in the miniMOM scheme \cite{vonSmekal:2009ae}, the running gauge coupling will just drop as given in equation \pref{alphar}, i.\ e.\ purely logarithmically.

In the same way also averaged tensor structures for the vertices can be constructed, for all possible globally invariant gauge tensor structures \cite{Macher:2011ad}, and in the same way as before related to the ones of 't Hooft gauge. Hence, the perturbative results can indeed be obtained relatively straightforwardly. Especially, only gauge algebra is required, and no new Feynman diagrams have to be evaluated. Thus, reliable statements about perturbative results in the main text are possible.

\bibliographystyle{bibstyle}
\bibliography{bib}


\end{document}